\documentclass[fleqn,usenatbib]{mnras}

\usepackage{newtxtext,newtxmath}

\usepackage[T1]{fontenc}
\usepackage{ae,aecompl}

\usepackage{graphicx} 
\usepackage{amsmath}  
\usepackage{amssymb}  
\usepackage{mathtools}
\usepackage{bm}

\usepackage{tikz}
  \usetikzlibrary{patterns, angles, quotes}
  \usetikzlibrary{decorations.pathreplacing}

  \pgfdeclarepatternformonly[\hatchspread,\hatchthickness,\hatchshift,\hatchcolor]
     {custom north east lines}
     {\pgfqpoint{\dimexpr-2\hatchthickness}{\dimexpr-2\hatchthickness}}
     {\pgfqpoint{\dimexpr\hatchspread+2\hatchthickness}{\dimexpr\hatchspread+2\hatchthickness}}
     {\pgfqpoint{\dimexpr\hatchspread}{\dimexpr\hatchspread}}
     {
      \pgfsetlinewidth{\hatchthickness}
      \pgfpathmoveto{\pgfqpoint{\dimexpr\hatchshift-0.15pt}{-0.15pt}}
      \pgfpathlineto{\pgfqpoint{\dimexpr\hatchspread+0.15pt}{\dimexpr\hatchspread-\hatchshift+0.15pt}}
      \ifdim \hatchshift > 0pt
        \pgfpathmoveto{\pgfqpoint{-0.15pt}{\dimexpr\hatchspread-\hatchshift-0.15pt}}
        \pgfpathlineto{\pgfqpoint{\dimexpr\hatchshift+0.15pt}{\dimexpr\hatchspread+0.15pt}}
      \fi
      \pgfsetstrokecolor{\hatchcolor}
      \pgfusepath{stroke}
     }
\usepackage{tikzscale}
\usepackage{tkz-euclide}
\usetkzobj{all}

\usepackage{etoolbox}
\makeatletter
\patchcmd\@combinedblfloats{\box\@outputbox}{\unvbox\@outputbox}{}{\errmessage{\noexpand patch failed}}
\makeatother

\title[Late X-ray flares from RS-ejecta interaction]{Late X-ray flares from the interaction of a reverse shock with a stratified ejecta in GRB afterglows: simulations on a moving mesh}

\author[E. H. Ayache et al.]{
Eliot H. Ayache,$^{1}$\thanks{E-mail: e.h.r.ayache@bath.ac.uk (EHA)}
Hendrik J. van Eerten,$^{1}$
Fr\'ed\'eric Daigne$^{2}$
\\
$^{1}$Department of Physics, University of Bath, Claverton Down, BA2 7AY, UK\\
$^{2}$Sorbonne Universit\'e, CNRS, UMR 7095, Institut d'Astrophysique de Paris, 98 bis Boulevard Arago, 75014 Paris, France
}

\date{Accepted 2020 May 18. Received 2020 May 18; in original form 2020 April 8}

\pubyear{2020}

\begin{document}
\label{firstpage}
\pagerange{\pageref{firstpage}--\pageref{lastpage}}
\maketitle

\begin{abstract}
  
  Late activity of the central engine is often invoked in order to explain the flares observed in the early X-ray afterglow of gamma-ray bursts, either in the form of an active neutron star remnant or (fall-back) accretion onto a black hole. However, these scenarios are not always plausible, in particular when flares are delayed to very late times after the burst. Recently, a new scenario was proposed that suggests X-ray flares can be the result of the passing of a long-lived reverse shock through a stratified ejecta, with the advantage that it does not require late-time engine activity.
  In this work, we numerically demonstrate this scenario to be physically plausible, by performing onedimensional simulations of ejecta dynamics and emission using our novel moving-mesh relativistic hydrodynamics code. Improved efficiency and precision over previous work enables the exploration of a broader range of setups. We can introduce a more physically realistic description of the circumburst medium mass density. We can also locally trace the cooling of electrons when computing the broadband emission from these setups. 
  We show that the synchrotron cooling timescale can dominate the flare decay time if the stratification in the ejecta is constrained to a localised angular region inside the jet, with size corresponding to the relativistic causal connection angle, and that it corresponds to values reported in observations. We demonstrate that this scenario can produce a large range of observed flare times, suggesting a connection between flares and initial ejection dynamics rather than with late-time remnant activity.

\end{abstract}




\section{Introduction}

Gamma-Ray Bursts (GRBs) are the most luminous explosions in the Universe. They are the result of the collapse of a massive star (long GRBs) \citep{Woosley1993,MacFadyen1999a} or of the merger of two Neutron Stars (NS) or a NS with a black hole (Short GRBs) \citep{Eichler1989,Mochkovitch1995}, a scenario spectacularly confirmed two years ago with the joint detection of GW170817 and GRB170817a \citep{Abbott2017b,Abbott2017,Goldstein2017,Savchenko2017,Troja2017a,Hallinan2017}. These explosions give rise to a collimated ultra-relativistic jetted outflow typically observed on-axis. In the first milliseconds to a few seconds, an initial burst of gamma-rays (the prompt emission) can be observed, followed after a few hundred seconds by an afterglow (at larger wavelengths). The afterglow, originally peaking at X-ray energies, decreases in luminosity and frequency with time until it disappears. 

The launch of the \textit{Neil Gehrels Swift} space observatory \citep{Gehrels2004} in 2004, dedicated to GRB detection and early X-ray/optical afterglow observations, showed the great variability in the light curve at early times. For 15 years now, \textit{Swift} has been detecting GRBs and measuring the properties and spectra of their early X-ray afterglow at a rate of around 90 detections a year \citep{Evans2009}, revealing flaring and re-brightening behaviors. These flares are detected in about a third of GRBs \citep{Burrows2005,Falcone2007,Chincarini2010,Margutti2011,Yi2016} and display a characteristic shape with a sharp jump in luminosity followed by a smoother decay, eventually going back to the original luminosity level \citep{Chincarini2007}. They can happen at any time during the afterglow, from $10^2$ to $10^5$ s after the prompt emission, growing longer with time following  $\Delta t_\mathrm{flare}/ t_\mathrm{flare} \sim 0.1 - 0.3$ \citep[e.g.][]{Burrows2005}. \citet{Margutti2011} showed that the average isotropic equivalent energy of flares also decreased with time as $E_\mathrm{iso, flare} \propto t_\mathrm{obs, flare}^{-1.7}$, albeit with a very large scatter.

Several works have shown that the external forward shock (FS) cannot produce flares with the correct timescale and flux variability \citep{Ioka2005,Nakar2007,VanEerten2009}, and the currently accepted paradigm involves late-time activity of the central engine \citep{Burrows2005,Chincarini2010,Margutti2011} or late-time fallback accretion onto the remnant \citep{King2005,Perna2006,Proga2006,Lee2009,Cao2014}. This scenario, though, requires the central engine to be active on the same timescale in the lab frame as that of the occurrence of the flare \citep{Kobayashi2005}, implying that the engine has not yet collapsed to a black hole in cases where fallback accretion is unlikely to play a role as well (e.g. at very late timescales). It also requires this surge of activity to obey the constraint $\Delta t / t \sim 0.1$, putting an intrinsinc constraint on the variability timescale of the activity of the central engine up to very late times.
Consequently, having a model that relaxes the constraints on the central engine is a step towards better understanding of late-time variability in X-ray afterglows.

In order to relieve the constraints on the central engine, \citet{Hascoet2017} proposed an alternate mechanism in which flares can arise from the interaction of a long-lived reverse shock (RS) with a stratified ejecta produced by a gradual and non-monotonic shutdown of the central engine right after the initial ejection phase, and tested this scenario using a ballistic model. Indeed, \citet{Uhm2007} and \citet{Genet2007a} showed that the emissivity of the RS can dominate over that of the FS. A dominating RS is also able to reproduce most of the other canonical features of the early X-ray afterglow \citep{Uhm2012,VanEerten2014}. \citet{Lamberts2017a} \defcitealias{Lamberts2017a}{LD18} (hereinafter \citetalias{Lamberts2017a}) further investigated this approach by running complete hydrodynamical simulations using the relativistic version of the adaptive mesh refinement (AMR) hydrodynamics code RAMSES \citep{Teyssier2002,Lamberts2013}. They were able to reproduce in principle flaring behavior and validate the ballistic model of  \citet{Hascoet2017}. But even with using a moving grid where cells were deleted behind the ejecta to be reused upstream, they were limited by the fixed resolution of their numerical hydrodynamics code. As a result they were constrained to very dense uniform cirumburst environment (density $n_0 \sim 10^3 \text{ cm}^{-3}$). Additionally, they were not able to directly trace the impact of synchrotron cooling on electron populations as fluid zones advect away from the shocks. Finally, the study was limited in the range of observer times for flares it could demonstrate.

In this work, we expand the approach of \citetalias{Lamberts2017a} to more realistic dynamical initial conditions. We use a newly developed moving mesh special relativistic hydrodynamics code that allows for improved numerical accuracy and computational efficiency. We improve the radiative prescription by implementing local cooling, by tracing the electron Lorentz factors bracketing the accelerated electron population through the fluid \citep{Downes2002,VanEerten2010}.

The hydrodynamics code we used and the simulation setup are described in section \ref{sec:numerical_methods}. The dynamical evolution is described in section \ref{sec:hydrodynamical_evolution}. We explain our emission prescription and discuss flare production in section \ref{sec:emission}.

\section{Numerical Methods}
\label{sec:numerical_methods}

\subsection{Relativistic hydrodynamics on a moving mesh}

The simulations were ran using our own numerical dynamics code (see appendix \ref{sec:code_validation} for code validation). 
Similarly to the codes from \citet{Kobayashi1999}, \citet{Daigne2000} and \citet{Duffell2011}, our code uses a special relativistic hydrodynamics (SRHD) finite-volume Godunov prescription on a moving mesh in spherical coordinates $(r, \theta, \varphi)$, solving the following system of conservation equations:
\begin{align}
    \partial_t \bm{U} + \bm{\nabla F}(\bm{U}) = \bm{S},
    \label{eq:conservativeFormulation}
\end{align}
where $\bm{U}$ and $\bm{F}(\bm{U})$ are the vector of conserved variables and the corresponding flux vector, respectively and $\bm{S}$ is the source term. $\bm{U}$ and $\bm{F}(\bm{U})$ are given by:
\begin{align}
    \bm{U} =
    \begin{pmatrix}
      D \\ m \\ E 
    \end{pmatrix}
    \equiv
    \begin{pmatrix}
      \rho \Gamma \\ \rho h \Gamma^2 v \\ \rho h \Gamma^2 - p
    \end{pmatrix}  &&        
    \begin{matrix*}[l]
      \text{(Rest-mass density)}\\
      \text{(Momentum)}\\
      \text{(Energy)}
    \end{matrix*},
\end{align}
\begin{align}
    \bm{F}(\bm{U}) = 
    \begin{pmatrix}
      D v \\ mv + p \\ m
    \end{pmatrix},
\label{eq:conserved_variables}
\end{align}
where the primitive variables $\rho, v, p$ are the rest-mass density, the fluid velocity and the fluid pressure in the comoving frame, respectively, $h$ is the specific enthalpy, $\Gamma$ the Lorentz factor of the fluid, and the speed of light $c = 1$. The spherical coordinates formulation of these equations includes a non-zero source term $\bm{S} = (0, 2 p/r, 0)^T$.

The system of equations is closed using an ideal monoatomic fluid equation of state:
\begin{align}
    p(\rho, \epsilon) = \rho \epsilon (\gamma - 1),
\end{align}
where $\epsilon$ is the specific internal energy density, and $\gamma$ the adiabatic index of the fluid set to $\gamma = 4/3$ in the ultra-relativistic case.

The 1D version of the code involves a grid of $N$ cells $(\mathcal{C}_i)_{i\in\llbracket0,N-1\rrbracket}$ along the $r$ direction, for which the cell-centered vector of conserved variables is written $\bm{U}_{i}$, and $N-1$ interfaces $(\mathcal{I}_j)_{j\in\llbracket1/2,N-1-1/2\rrbracket}$ between these cells. As in any Godunov approach, computing the evolution of the fluid involves determining the fluxes at each interface by solving a Riemann problem at a given time. The code is built on an Arbitrary Lagrangian-Eulerian (ALE) approach where the interfaces are allowed to move at arbitrary velocity (provided that it does not exceed the speed of light $c$). We set this velocity to the velocity of the contact discontinuity in the Riemann fan at each interface. This makes the code purely Lagrangian. The fluxes are then computed by solving the Riemann problem in the frame of reference of the interface. We then transform these fluxes to the lab frame before updating the cell averages.

We use a HLLC Riemann solver \citep{Harten1983,Mignone&Bodo2006} and linear spatial reconstruction with a minmod slope limiter. Time integration is done using a third order Runge-Kutta prescription which has the advantage of being total variation diminishing, with a Courant-Friedrich-Lewy condition of 0.2.

The moving-mesh approach introduces the need for re-gridding. Cells that become too small will hinder computational efficiency by driving down the time-step, while we might lose crucial resolution in other areas because of expanding cells. The former issue is addressed by merging neighbouring cells while the latter is solved by splitting cells following a linear interpolation with a minmod slope limiter to ensure conservation. Re-gridding criteria are twofold: first, a maximum and minimum cell aspect ratio constrain resolution for a given radius (a cell gets refined when it's aspect ratio exceeds the upper bound, and de-refined when it drops below the lower bound); second, we enforce boundaries on primitive variables spatial second derivatives in order to increase the resolution of regions with small-scale variability. Mesh regularity is enforced by splitting any cell becoming more than three times as big as one of its neighbours, provided that the two previous criteria are respected. All these parameters can be updated on the fly when experimenting with new setups, by taking advantage of the code to start from any previous data dump. The re-gridding parameters we used in our simulations are described in section \ref{sub:simulation_setup}.

Computational efficiency in the GRB setup is further improved by introducing a moving outer boundary traveling at the speed of light ahead of the forward shock. The rightmost cell $\mathcal{C}_{N-1}$ is split if $\Delta r_{N-1} > 2 \Delta r_0 \times \frac{r_{N-1}}{r_{N-1}(t=0)}$, thus resulting in a dynamically  produced logarithmic grid ahead of the FS.

We assess the code accuracy and stability on standard test setups for which we are able to compute the exact solution and show that we achieve second order convergence for smooth flows while conserving accurate shock descriptions and evolution. The tests and the associated results are described in detail in appendix~\ref{sec:code_validation}. As a result, we can confidently use the results of the dynamical simulations in more elaborate cases. 

The computation of radiative emission relies on shock detection in the dynamics of the blast-wave. We choose to use a modified version of the shock detector used in \citetalias{Lamberts2017a} based on the method introduced in \citet{Rezzolla2003} and described in \citet{Zanotti2010}, that evaluates the nature of the waves in the Riemann fan for all interfaces of the grid based on the calculation of limiting relative velocities $\tilde{v}_{12}$. These waves can either be rarefaction waves or shocks. Physical shocks will be signaled by shocks waves emerging from these interfaces with relative velocity $v_{12} > \tilde{v}_{12}$. In practice, we place shocks at the locations of local maxima of $S = v_{12} - \tilde{v}_{12}$. This method, however, can lead to the spurious detection of weak shocks. \citetalias{Lamberts2017a} address this issue by smoothing $S$. The intense re-gridding used in our approach introduces occasional additional spurious variability in the value of S for which the smoothing approach is unsatisfying. However, the 1D aspect of our simulation makes the implementation of the full method from \citet{Rezzolla2003} rather straightforward. This method, which we implemented successfully, tightens the shock detection threshold by also taking into account the limiting relative velocity corresponding to the 2-shock Riemann fan. The calculation of the limiting relative velocities is described in appendix~\ref{sec:shock_detection_algorithm}, along with the treatment of weak shocks.

\subsection{Simulation setup}
\label{sub:simulation_setup}

  Our approach follows that of \citetalias{Lamberts2017a}, that we describe in this section before introducing the improvements we carried out. We emulate the behavior of the central engine during ejection by setting up a slab of ejecta, shortly after breakout, in which we set up a variable initial radial velocity profile. We will consider a constant power of the central engine during the ejection phase, meaning that faster material will also be more diffuse. A gradual shutdown can then be represented by slower material towards the back of the ejecta. The chaotic aspect of that shutdown is tackled by introducing non-monotone radial variability of the velocity in this region of the fluid. 

  In this approach, the initial state is characterized by a cold ejecta with variable Lorentz factor radial profile $\Gamma(r)$. The total isotropic equivalent energy of the blast wave is set to $E = 10^{53} \text{ erg}$ and the initial ejection happens between $t_0 = 0 \text{ s}$ and $t_w = 100 \text{ s}$, bringing the constant isotropic power of the ejection to $\dot{E} = 10^{51} \text{ erg}.\mathrm{s}^{-1}$. The ejecta being cold, the thermal pressure $p$ is set by $\eta = p/\rho c^2 = 10^{-3}$, where the rest-mass density profile is expressed as a function of $\dot{E}$ and $\Gamma(r)$ as follows:
  \begin{align}
    \rho (r) = \frac{\dot{E}}{4 \pi r^2 v(r) \Gamma^2(r) c^2 \left[ 1 + \eta \left( \frac{\gamma}{ \gamma - 1} - \frac{1}{\Gamma^2(r)}\right)\right]},
  \end{align}
  where $v(r)$ is the velocity of the fluid. 

  Until the starting time of the dynamical simulation $t_\mathrm{ini} = 200 \text{s}$, the ejecta is assumed to follow a ballistic evolution, allowing to compute the corresponding positions of the external and internal edges of the ejected shell, respectively $R_0$ and $R_\Delta$ (with Lorentz factors $\Gamma_0$ and $\Gamma_\Delta$), such that:
  \begin{align}
    R_0 &= \left(1 - \frac{1}{\Gamma_0^2}\right)^{1/2} c (t_\mathrm{ini} - t_0),\\
    R_\Delta &= \left(1 - \frac{1}{\Gamma_\Delta^2}\right)^{1/2} c (t_\mathrm{ini} - t_w - t_0).
  \end{align}
  Since $\dot{E}$ is set to a constant value and the ejecta is cold, the initial radial profile can be fully described by a single physical quantity. Here we follow \citetalias{Lamberts2017a} and set the value of the Lorentz factor $\Gamma(r)$. The shell will be characterized by an outer flat head region with Lorentz factor $\Gamma_0$ and a region of decreasing velocity (the tail region) all the way to engine shutdown at Lorentz factor $\Gamma_\Delta$. The initial Lorentz factor profile is given by:
  \begin{align}
    \Gamma(r) = 
    \begin{cases}
      1 & \text{ if $0 < r \leq R_\Delta $}, \\
      [\Gamma_\Delta + (\Gamma_0 - \Gamma_\Delta) x][1 + f(x)]
          & \text{ if $R_\Delta < r \leq r_\alpha $}, \\
      \Gamma_0      & \text{ if $r_\alpha < r \leq R_0 $}, \\
      1             & \text{ if $R_0 < r$},
    \end{cases}
    \label{eq:initial_gamma_profile}
  \end{align}
  with,
  \begin{align}
    x = \frac{r - R_\Delta}{r_\alpha - R_\Delta},
  \end{align}
  where $r_\alpha = R_\Delta + \alpha(R_0 - R_\Delta)$ sets the limit between the head and tail regions of the ejecta ($\alpha$ is the ratio of tail to head). $f(x)$ is the profile of the perturbation responsible for the stratification of the ejecta and emanating from a chaotic shutdown of the central engine. In \citetalias{Lamberts2017a}, $f$ is given by:
  \begin{align}
    f(x) = A \sin(2\pi x), && A \in [0,1],
  \end{align}
  where A is the amplitude of the perturbation.

\begin{figure}
    \centering
    \includegraphics[width=0.4\textwidth]{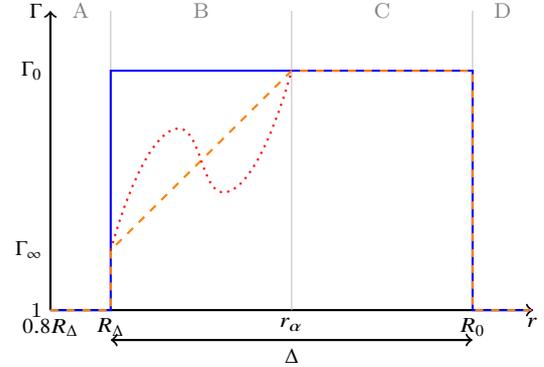}
    \caption{Schematic representation of the initial profile of the Lorentz factor $\Gamma$ as a function of the radius $r$. The solid blue line represents the \emph{pulse} profile with constant $\Gamma$ throughout all the ejection. The dashed orange line represents the case with a tail of Lorentz factor $\Gamma$ (progressive shutdown of the central engine). The dotted red line is the case with a non-monotonic progressive shutdown, expected to lead to the formation of internal shocks in the tail.}
    \label{fig:lfac_ini}
  \end{figure}

  We run SRHD simulations for various values of $A$ and $\alpha$ (reported in table~\ref{tab:run_parameters}), with the corresponding profiles drawn in figure \ref{fig:lfac_ini}. These values are the same as in \citetalias{Lamberts2017a} to allow for comparison. Run1 is a test of our setup for a canonical reverse shock (RS) propagating inside a homogeneous shell of ejecta. Run2 simulates the propagation of a long-lived RS (LLRS) with a significant "tail" in the ejecta, corresponding to the gradual shutdown of the central engine. Run3 produces a flare as a result of its non-monotonic tail (corresponding to non-monotonic central engine shutdown) that produces stratification in the ejecta. The initial profile parameters are summarized in table \ref{tab:initial_parameters}. 

  The grid has outflow boundary conditions on both sides. The choice of AMR criteria is driven by both the need for high resolution downstream of the shocks, as we expect very short synchrotron cooling timescales (see sec. \ref{sub:radiative_prescription}), and by the need for computational efficiency. These criteria will be different for the ejecta, the external medium, and the material behind the ejecta. For this reason we track different regions of the fluid using a passive scalar. The upper and lower limits on aspect ratio for each part of the fluid are reported in table \ref{tab:AMR}. We find a good balance between FS resolution and efficiency for a higher bound decreasing with $r$ in the circumburst medium and adjust the power-law index manually. We also add in this part of the flow a de-refinement criterion based on the derivative of the synchrotron cooling rate downstream of the FS to improve efficiency that we also adjust manually.

  We run simulations starting with 1200 cells in the ejecta and 100 in the external medium. Depending on the setup, the total number of cells at the end of the simulations is usually contained between 3000 and 5000.

  \begin{table}
    \small
    \begin{tabular}{lrrl}
      Run & $\alpha$ & $A$ & Description \\
      \hline
      run1 & 0   & 0 & homogeneous\\
      run2 & 0.5 & 0 & tail\\
      run3 & 0.5 & 0.6 & perturbed\\
      \hline
    \end{tabular}
    \caption{Run Parameters}
    \label{tab:run_parameters}
  \end{table}

  \begin{table}
    \small
    \begin{tabular}{llrl}
      \hline
      \textbf{Parameter} & \textbf{Notation} & \textbf{Value} & \textbf{Unit}\\
      \hline
      Injected kinetic power  &  $\dot{E}$  & $10^{51}$ & erg/s \\
      Burst Lorentz Factor    &  $\Gamma_0$ & $100$     &       \\
      Tail Lorentz Factor     &  $\Gamma_\Delta$ & $10$      &       \\
      Burst starting time     &  $t_0$      & $ 0$      & s     \\
      Duration of ejection    &  $t_w$      & $100$     & s     \\
      Initial time of simulation & $t_\mathrm{ini}$ & $200$ & s \\
      External medium number density  & $n_0$   & $1$   & $\mathrm{cm}^{-3}$\\
      Ratio between pressure $p$ and $\rho c^2$ &  $\eta$ & $10^{-3}$ & \\
      \hline
    \end{tabular}
    \caption{Initial parameters}
    \label{tab:initial_parameters}
  \end{table}

  \begin{table}
    \small
    \begin{tabular}{lll}
      Fluid material & Lower bound & Higher bound \\
      \hline
      ejecta             & $10^{-10}$ & $5 \times 10^{-4}$ \\
      circumburst medium & $10^{-9}$ & $5 \times 10^{-5} \times (r/R_0)^{-0.1}$ \\
      \hline
    \end{tabular}
    \caption{Adaptive mesh refinement aspect ratio criteria}
    \label{tab:AMR}
  \end{table}

\section{Hydrodynamical evolution}
\label{sec:hydrodynamical_evolution}

    In this section, we comment on the behavior of the fluid resulting from the various setups we introduced. The results are in accordance with \citetalias{Lamberts2017a} and we refer the reader to this paper for a more detailed discussion of the dynamics. 

\begin{figure}
      \centering
      \includegraphics[width=0.52\textwidth]{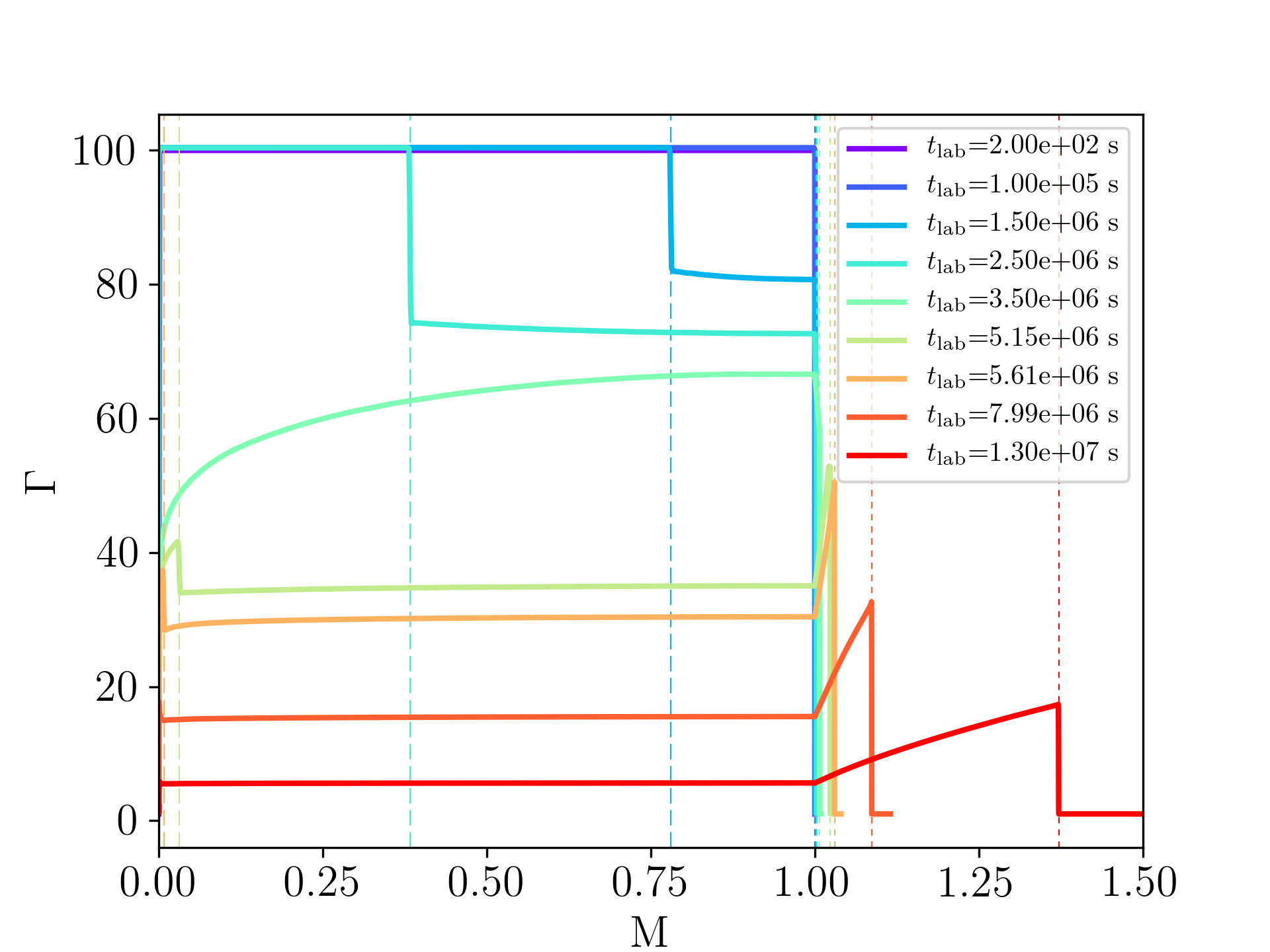}
      \includegraphics[width=0.52\textwidth]{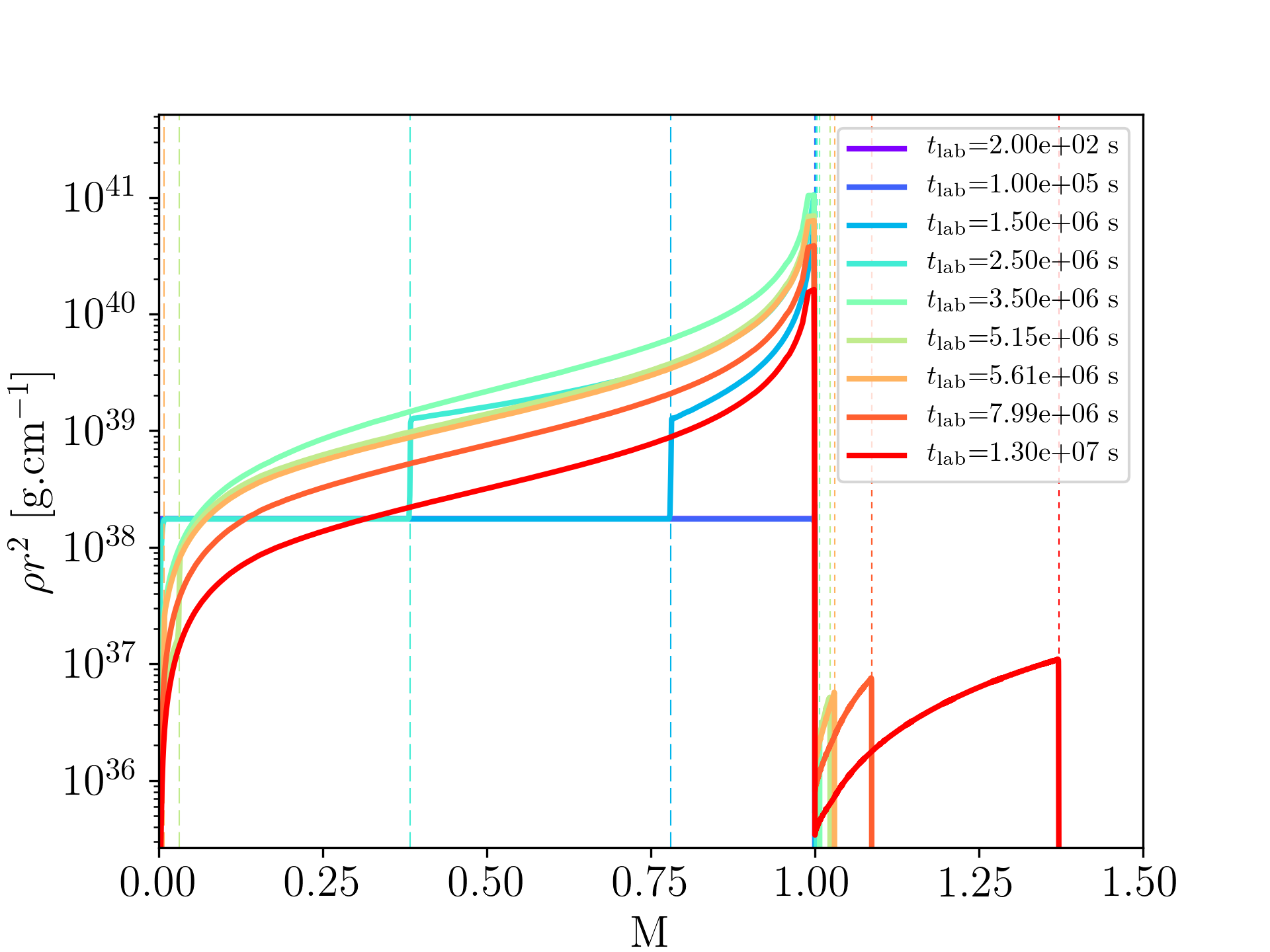}
      \caption{Lorentz factor (top) and density (bottom) as a function of cumulative mass for run1 (homogeneous ejecta). Dashed vertical lines show the positions of reverse shocks and dotted vertical lines the positions of forward shocks.}
      \label{fig:run1_dynamics}
    \end{figure}

    The dynamical evolution of run1 ("homogeneous shell") is reported in fig.~\ref{fig:run1_dynamics}. In order to make dynamical evolution figures more easily readable, we plot physical quantities as a function of cumulated mass $M$ defined as follows:
    \begin{align}
      M(r) = \frac{1}{M_0} \int_{R_\Delta}^{r} 4 \pi \rho \Gamma (r^\prime)^2 \mathrm{d}r^\prime
    \end{align}
    As expected, one can observe the formation of a reverse shock propagating leftwards in the fluid frame (blue curve at $1.5 \times 10^{5}$ s), with a crossing time of $t_\mathrm{cross} < 3.50 \times 10^{6}$s. In their simulations, \citeauthor{Lamberts2017a} were limited in the spatial scale they could cover and thus discarded the tail of the ejecta as the shocked medium grew in mass so as to keep a constant total simulated mass. Thanks to our numerical prescription, we are able to keep all the initial material in the simulation. This additional material at the back of the ejecta is responsible for the formation of a spurious internal reverse shock, visible here at $M \sim 0.05$, $t_\mathrm{lab} = 5.15\times 10^6$ s (green curve), after the external reverse shock has crossed the ejecta.
    After deceleration, a forward shock starts to form ahead of the ejecta (M > 1) as the blast wave sweeps up some circumburst medium material.

\begin{figure}
      \centering
      \includegraphics[width=0.52\textwidth]{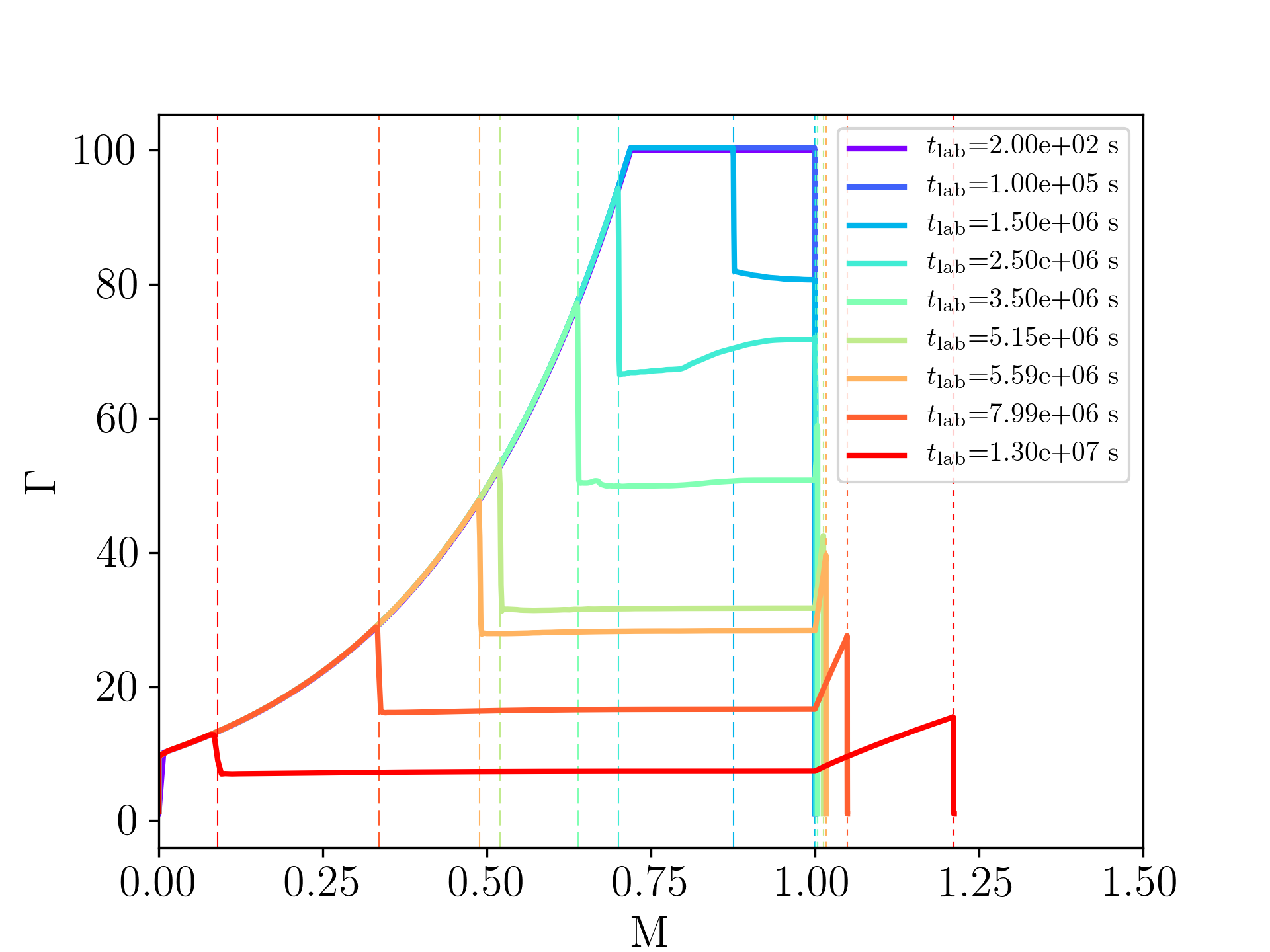}
      \includegraphics[width=0.52\textwidth]{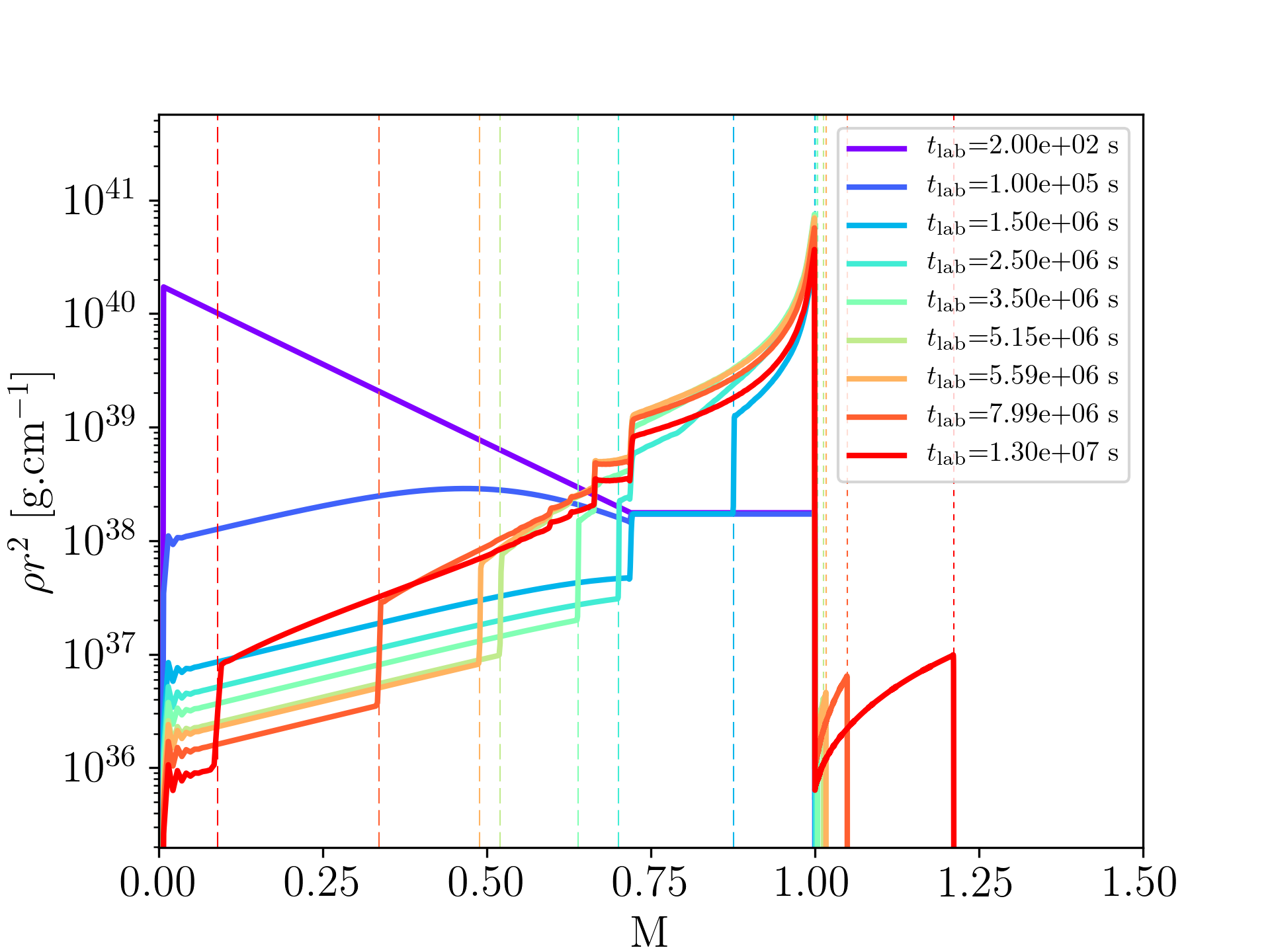}
      \caption{Lorentz factor (top) and density (bottom) as a function of cumulative mass for run2 (gradual shut down of the central engine leading to a region of decreasing Lorentz factor at the back of the ejecta). Dashed vertical lines show the positions of reverse shocks and dotted vertical lines the positions of forward shocks.}
      \label{fig:run2_dynamics}
    \end{figure}

    Fig.~\ref{fig:run2_dynamics} shows the dynamical results for run2 ("decreasing tail"). The initial profile of velocity increasing with $r$ is responsible for the tail expanding, as is visible in the density profile (lower pane) where the density decreases in the tail ($M \lesssim 0.7$) as the blast wave moves to larger radii. This causes the RS to take longer to traverse the ejecta, thus giving rise to a long-lived reverse shock. Indeed, the crossing time is increased to $t_\mathrm{cross} > 1.3 \times 10^{7}$ s in this case, thus giving rise to a long-lived RS.

\begin{figure}
      \centering
      \includegraphics[width=0.52\textwidth]{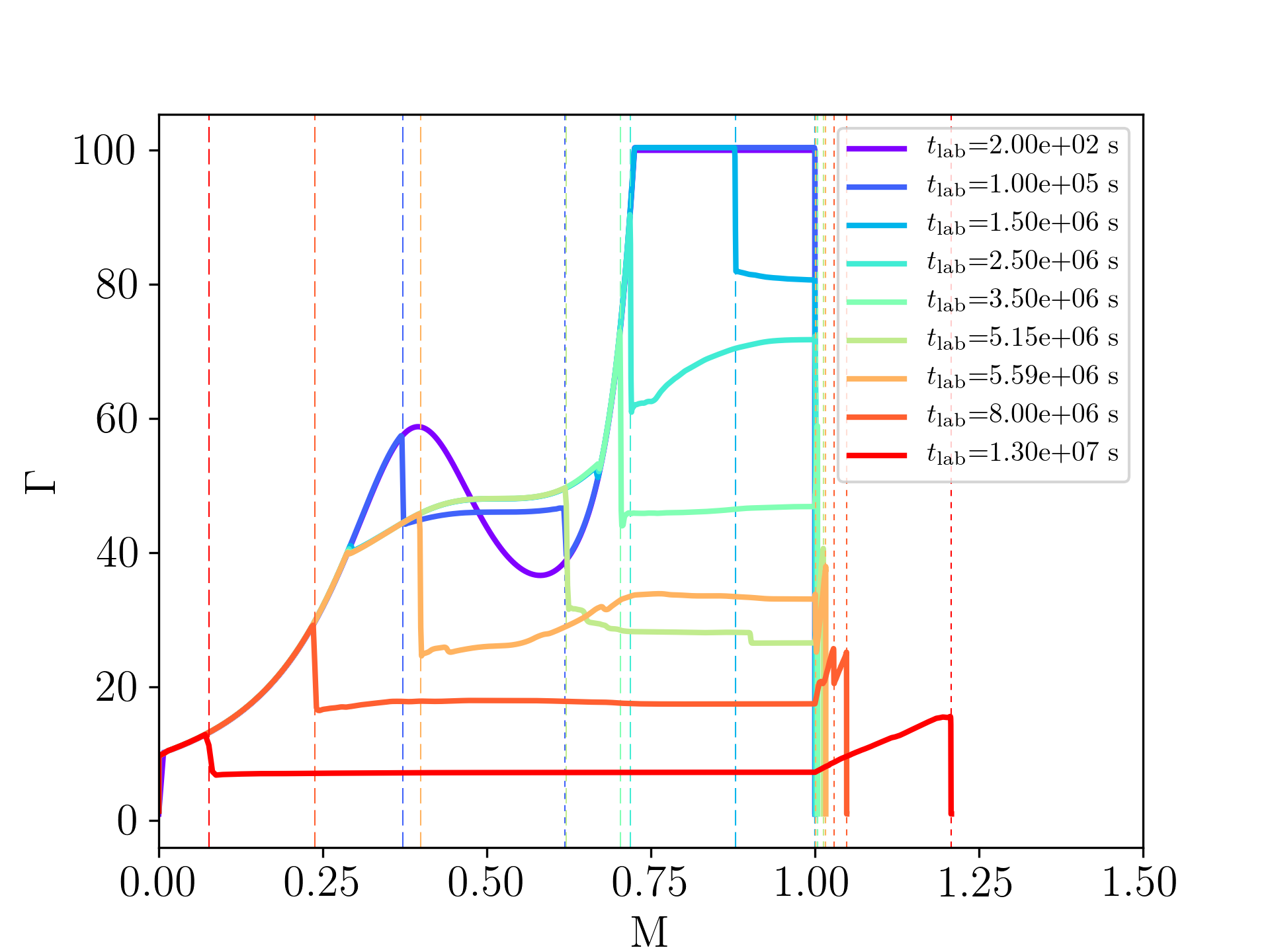}
      \includegraphics[width=0.52\textwidth]{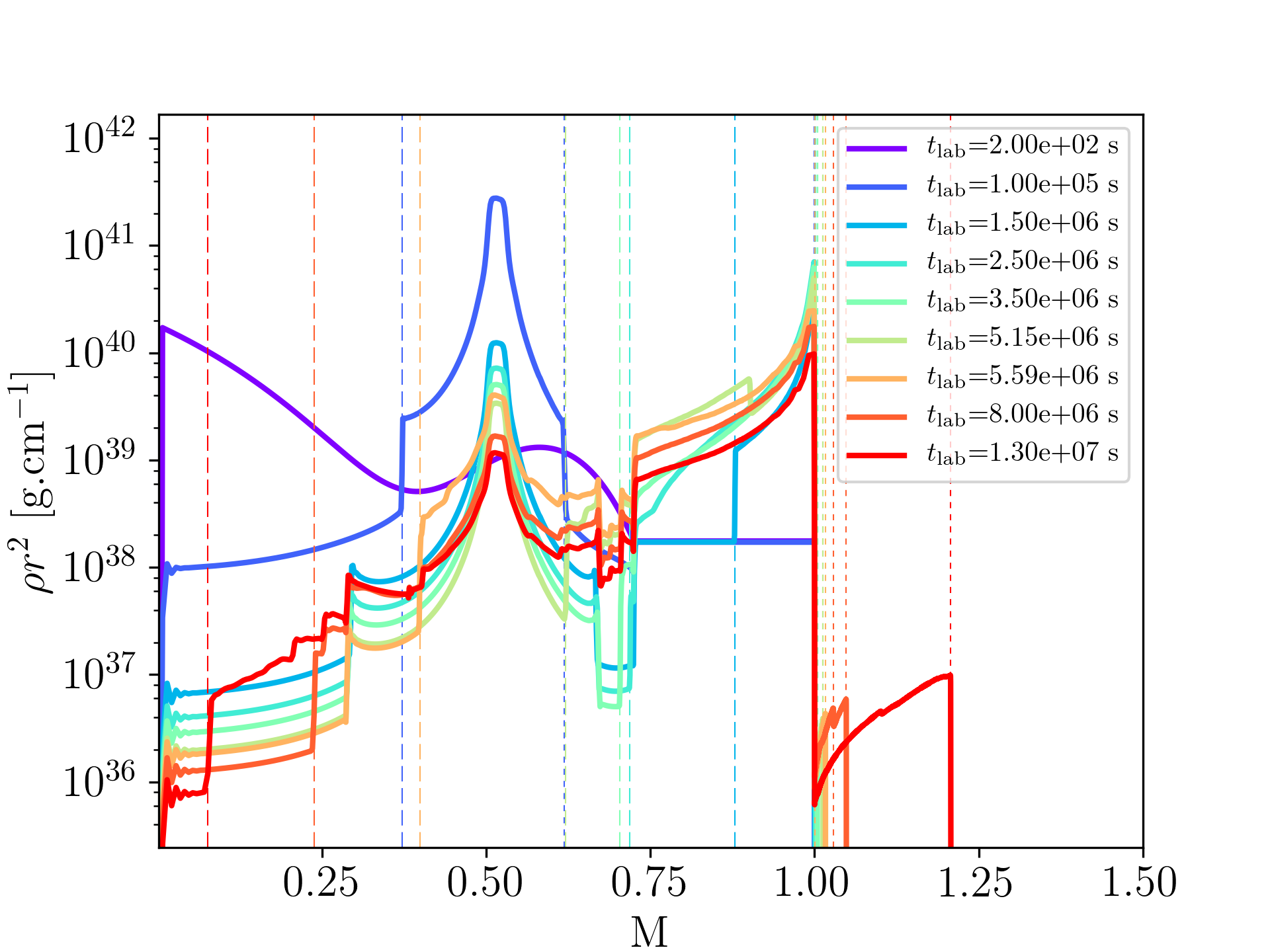}
      \caption{Lorentz factor (top) and density (bottom) as a function of cumulative mass for run3 (chaotic shut down of the central engine leading to a region of negative velocity gradient in the trail of the ejecta. Dashed vertical lines show the positions of reverse shocks and dotted vertical lines the positions of forward shocks.)}
      \label{fig:run3_dynamics}
    \end{figure}

    The results of run3 are shown in fig.~\ref{fig:run3_dynamics}. The important feature in this run is the presence of a non-monotonic radial Lorentz factor profile in the tail of the ejecta, akin to the ballistic simulations of \citet{Hascoet2017}, and run 4 in \citetalias{Lamberts2017a}. The faster material at smaller radii eventually catches up with the slower material ahead, creating a first internal forward shock / reverse shock system (IFS/IRS) that delimits the boundaries of a dense shell of ejecta, with significantly higher rest-mass density (dark blue curve at $t_\mathrm{lab} = 1.00 \times 10^5$ s). The shocked material displays a centered spike in density that delimits the position of the contact discontinuity between "forward-shocked" and "reverse-shocked" material. This spike is not visible in the simulations of \citetalias{Lamberts2017a}, probably because of a lack of sufficient resolution, and is the result of the compression of the fluid ahead of shock formation because of the smooth velocity gradient in the initial setup. A similar feature has been predicted by numerical Lagrangian simulations and analytical calculations \citep[respectively]{Daigne2000,VanEerten2014a} and we can recover it here thanks to the moving-mesh approach. As the external RS enters the dense shell, it is partially reflected into a forward shock (light green curve, $t_\mathrm{lab} = 5.15 \times 10^6$ s) that eventually catches up with the external FS (yellow curve, $t_\mathrm{lab} = 5.59 \times 10^{6}$ s). As the RS progresses through the dense shell, it eventually interacts with the contact discontinuity density spike leading to the formation of an additional reflected forward shock that will eventually catch up with the previous one and strengthen it. We show however in section \ref{sub:emission_from_the_shocked_external_medium} that this feature does not influence the general shape of the light-curve and thus validate the results from \citetalias{Lamberts2017a}. Subsequently, the RS keeps traveling through the dense shell until it eventually leaves it before crossing the ejecta altogether for $t_\mathrm{cross} > 1.3 \times 10^7$ s. 

    All three setups display dynamical features similar to \citetalias{Lamberts2017a}, thus providing additional support to the idea that flares can be produced from the interaction of a LLRS with a stratified ejecta. Section \ref{sec:emission} focuses on the radiative prescription necessary to produce synthetic light-curves, and shows how run3 can be responsible for flaring behavior in GRB afterglows, while providing a better understanding of the spectral evolution of flares associated with this setup, thanks to the local cooling approach we implemented.


\section{Emission - Flares from a LLRS in a stratified ejecta}
\label{sec:emission}

\subsection{Radiative prescription - local cooling} 
\label{sub:radiative_prescription}

  Afterglow emission can be modeled by synchrotron radiation from electrons accelerated in shocks. In our situation, we can neglect self-absorption as we are only interested in the X-ray and optical afterglow. We ignore inverse Compton scattering for the sake of simplicity. We approximate the population of accelerated electrons by a truncated power-law:
  \begin{align}
    n^{\prime}(\gamma_e^{\prime}) = 
    \begin{cases}
      (\gamma_e^{\prime})^{-p} & \text{ if } \gamma_m < \gamma_e^{\prime} < \gamma_M ,\\
      0 & \text{ otherwise},
    \end{cases}
  \end{align}
  with $p = 2.3$ and $\gamma_m$ the energy of the bulk of the population, and $\gamma_M$ the evolving cut-off due to localised electron cooling, where the $^\prime$ notation denotes quantities in the fluid frame. We consider that a fraction $\zeta = 0.1$ of electrons present in the fluid are accelerated by any shock and that a fraction $\epsilon_e = 0.1$ of the internal energy density $e = \rho \epsilon$ is transferred to these electrons. Similarly, a fraction $\epsilon_B = 0.1$ of the internal energy density powers the tangled magnetic field needed for synchrotron radiation. 
  Electron cooling is driven by the adiabatic expansion and synchrotron cooling as shown by the following equation:
  \begin{align} 
    \label{eq:cooling}
    \frac{\mathrm{d}\gamma_e^\prime}{\mathrm{d}t^\prime} &=
      - \frac{\sigma_T (B^\prime)^2}{6 \pi m_e c } (\gamma_e^\prime)^2
      + \frac{\gamma_e^\prime}{3 \rho}\frac{\mathrm{d}\rho}{\mathrm{d}t^\prime},
  \end{align}
  with $\sigma_T$ the Thomson cross-section, $m_e$ the electron mass and $B^\prime = \sqrt{8 \pi \epsilon_B e}$ the magnetic field intensity. $\gamma_m$ and $\gamma_M$ follow this evolution. Since, our setups involve fast-cooling populations of electrons for which synchrotron losses need to be taken into account, we cannot compute $\gamma_m$ a posteriori and we apply the same method for the calculation of the local evolution of $\gamma_m$ and $\gamma_M$. Most works implement the global cooling approximation \citep{Granot2001,VanEerten2010a} that is based on assumptions on the dynamics downstream of the shock, or assume a single uniform electron population across the shocked fluid. The complex dynamics of our system prevents us from using this approximation as some regions of the fluid will be shocked multiple times. In their work, \citetalias{Lamberts2017a} approximate the cooling frequency by assuming it decreases with the dynamical timescale in the co-moving frame of the shocked region which approximates the time since a fluid parcel has been shocked. Our improved dynamical numerical scheme actually allows us to directly trace the local numerical evolution of $\gamma_m$ and $\gamma_M$, as long as we allow the resolution to increase sufficiently downstream of the shocks thanks to the compression of the mesh. Indeed, equation \ref{eq:cooling} can be rewritten in the following advection equation form \citep{Downes2002,VanEerten2010}:
  \begin{align}
    \label{eq:advect_cool}
    \frac{\partial}{\partial t} \left( \frac{\Gamma \rho^{4/3}}{\gamma_e^\prime} \right) + \frac{\partial}{\partial x^{i}} \left( \frac{\Gamma \rho^{4/3}}{\gamma_e^\prime} v \right)
      = \frac{\sigma_T}{6 \pi m_e c} \rho^{4/3} (B^\prime)^{2}.
  \end{align}
  The two boundaries of the electron population are thus simply handled as tracers during the hydrodynamical evolution. We reset their value every time a cell is shocked. In theory, the value of $\gamma_M$ directly downstream of a shock set by the acceleration timescale and can be taken to be infinity for sufficiently large p. In practice, we set $\gamma_M$ to $10^8$ when crossed by a shock, such that the induced frequency cutoff for the radiation from each individual shock is placed above 1keV in the observer frame. $\gamma_m$ is initialised by normalising the integrated energy of the electron population by the total available energy.

  The improved resolution downstream of the shocks, along with the fact that grid zones are co-moving with the fluid lift the constraints linked to the very short cooling time of the accelerated electrons and enables the implementation of local cooling with accurate treatment of new source term on the right-hand side of eq. \ref{eq:advect_cool}. 

\begin{figure}
    \centering
    \includegraphics[width=0.49\textwidth]{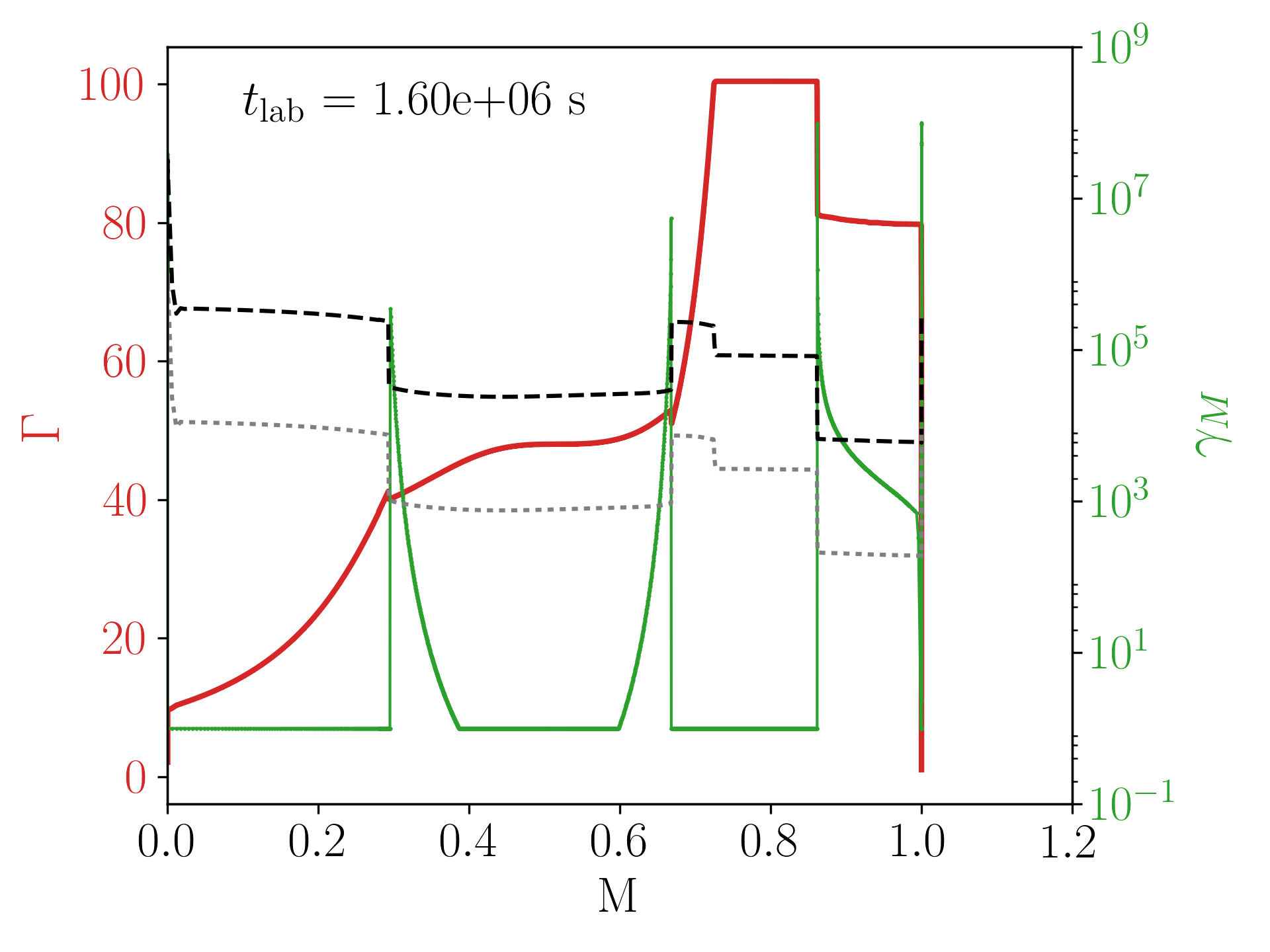}
    \includegraphics[width=0.49\textwidth]{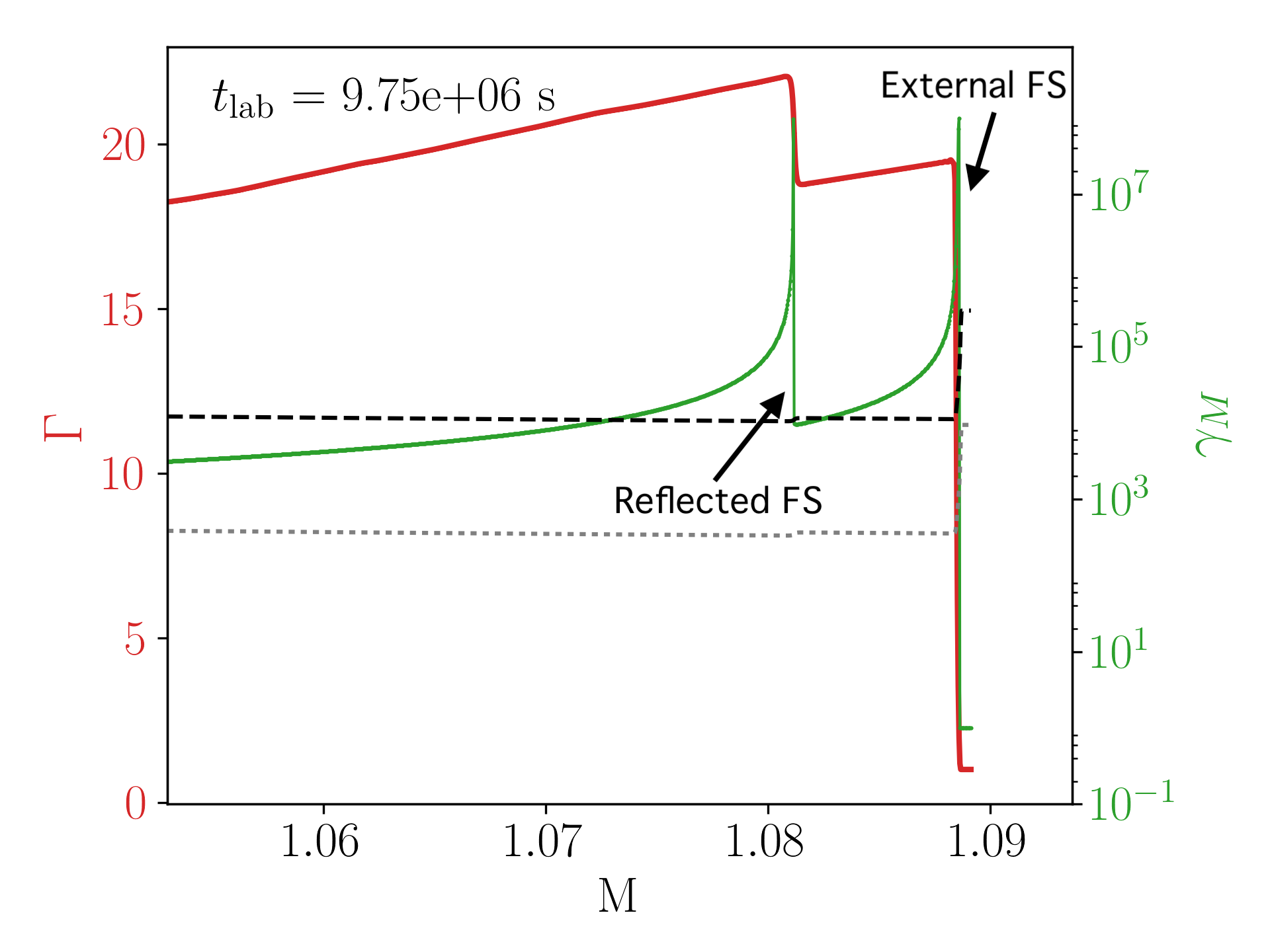}
    \caption{Snapshot of run3 at $t_\mathrm{lab}=1.60\times 10^{6}$ s (top panel) and $t_\mathrm{lab}=9.75\times 10^{6}$ s (bottom panel) showing the Lorentz factor (red) and $\gamma_M$ (green) as a function of cumulative mass $M$. In this figure $\epsilon_{B,\mathrm{FS}} = 0.1$. The black (grey) dashed line is the minimum value of $\gamma_M$ for which the synchrotron emission can contribute to X-rays (optical) in the observer frame. The bottom panel is zoomed in on the external medium to show the discrepancy between dynamical and synchrotron cooling timescale.}
    \label{fig:local_cooling}
  \end{figure}

  Figure \ref{fig:local_cooling} shows the values of $\gamma_M$ as a function of $M$ during the propagation of the reverse shock in the ejecta for run3 (perturbed setup). To begin with, the profile shows a complex pattern of multiple acceleration sites that cannot be modeled with a simple analytical approach. Second, $\gamma_M$ drops by several orders of magnitude over a very small number of cells downstream of each shock formed in the setup. This highlights the need for very high resolution, provided by the moving mesh, downstream of shocks for accurate calculation of the evolution of $\gamma_M$. Additionally, because of this very steep drop, the region contributing to X-ray emission downstream of the shocks is extremely narrow, adding an additional constraint on resolution in order to reconstruct the light-curve accurately. The region contributing to optical is wider, which is responsible for a smoother variations of the flux than in X-ray.

  Once the energy distribution of electrons is known, we can compute the spectral volumetric emitted power $P_\nu^\prime$ for any region of the flow downstream of a shock. Let $\nu_\mathrm{syn}(\gamma_e^\prime) = \frac{3q_e B^\prime}{16 m_e c} (\gamma_e^\prime)^2$ be the frequency of the synchrotron radiation from a single electron with energy $\gamma_e^\prime m_e c^2$ in the frame of the fluid, with $q_e$ the charge of the electron. We can associate a synchrotron frequency to any $\gamma_e$ such that $\nu_m^\prime = \nu_\mathrm{syn}(\gamma_m^\prime)$ and $\nu_M^\prime = \nu_\mathrm{syn}(\gamma_M^\prime)$. $P_\nu^\prime$ is thus given by:
  \begin{align}
    P_\nu^\prime = 
    \begin{cases}
      P_{\nu,\mathrm{max}}^\prime \left( \frac{\nu}{\nu_m^\prime} \right)^{1/3}, & \nu < \nu_m^\prime, \\
      P_{\nu,\mathrm{max}}^\prime \left( \frac{\nu}{\nu_m^\prime} \right)^{-(p-1)/2}, & \nu > \nu_m^\prime, 
    \end{cases}
  \end{align}
  \begin{align}
    &\text{with~} P_{\nu,\mathrm{max}}^\prime = \frac{4(p-1)}{3p - 1} 
      \times n^\prime \sigma_T \frac{4}{3} \frac{B^\prime}{6\pi} \frac{16m_e c}{3q_e}.
  \end{align}
  Above the maximum frequency $\nu_M^\prime$ the emitted power follows an exponential cut-off. We set $P_{\nu}^\prime = 0$ if $\nu > \nu_M^\prime$ for the sake of simplicity.

  We then integrate this power over the whole fluid using equation:
  \begin{align}
    F(\nu, t_\mathrm{obs}) = \frac{1+z}{2d_L^2}\int_{-1}^1\mathrm{d}\mu\int_0^\infty r^2 \mathrm{d}r \frac{P^\prime_{\nu^\prime}(r, t_\mathrm{obs}+r\mu)}{\Gamma^2 (1 - \beta\mu)^2}.
  \end{align}
  This integration assumes that the fluid is optically thin, a reasonable assumption above radio frequencies. For every cell in the volume, and a given lab time, we compute the arrival time for a redshift $z=1$, where we place the observer along the axis of the jet. The latter assumption allows to generalise the results to an observer within the opening angle of the flow, with the only difference being the nature of the jet break transition of the underlying non-flaring continuum. In practice, and since we only run 1D simulations, we divide the jet opening angle in 500 angular directions and compute the arrival times and radiative contribution for each cell in each one of these directions. The contributions from all cells for a given observer time bin are then summed to build the light-curve. As a sanity check, We verify a posteriori that we have taken into account the contributions from all the lab times corresponding to a given observer time by plotting the contribution to the total final light-curve for every snapshot.


\subsection{Synthetic light-curves and spectra} 
\label{sub:synthetic_light_curves_and_spectra}

\begin{figure}
      \centering
      \includegraphics[width=0.49\textwidth]{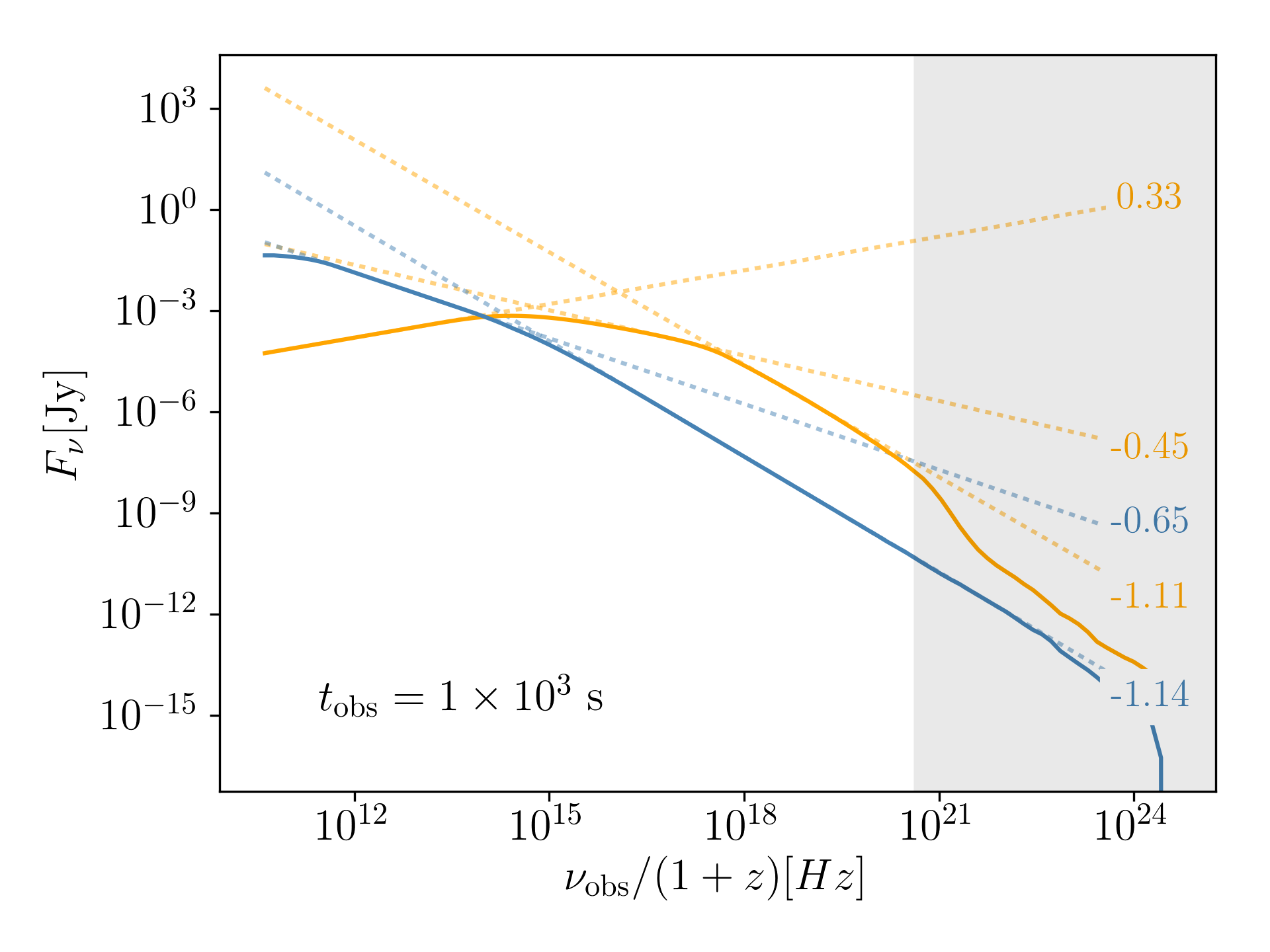}
      \caption{Spectral energy distribution for run1 (homogeneous) for the FS (orange) and the RS (blue). We fit each regime of these SEDs in log space and report the slope with the dotted lines. The choice of $\gamma_M$ directly downstream of the shocks causes the numerical solution to deviate from the expected slope at high energies beyond those that we model (shaded region).}
      \label{fig:spec_lim}
    \end{figure}

    Figure \ref{fig:spec_lim} shows the spectral energy distribution (SED) for the forward and the reverse shock for run1 (homogeneous).The slopes are in good agreement with analytical modeling of shock emission \citep{Sari1997a} and thus validate our radiative prescription. The slight deviation for the slopes of the FS decreases with increasing resolution downstream of the shock, at the expense of computational efficiency. We also confirm that the frequency cut-off resulting from our choice of initial value for $\gamma_M$ is also located above $10^{20}~\mathrm{Hz}$ for the FS and $10^{24}~\mathrm{Hz}$ for the RS. With observer time, this frequency cut-off decreases but we check that it always remains above X-ray frequencies, confirming the validity of our approach for light-curve calculations at 1keV.

\begin{figure}
      \centering
      \includegraphics[width=0.49\textwidth]{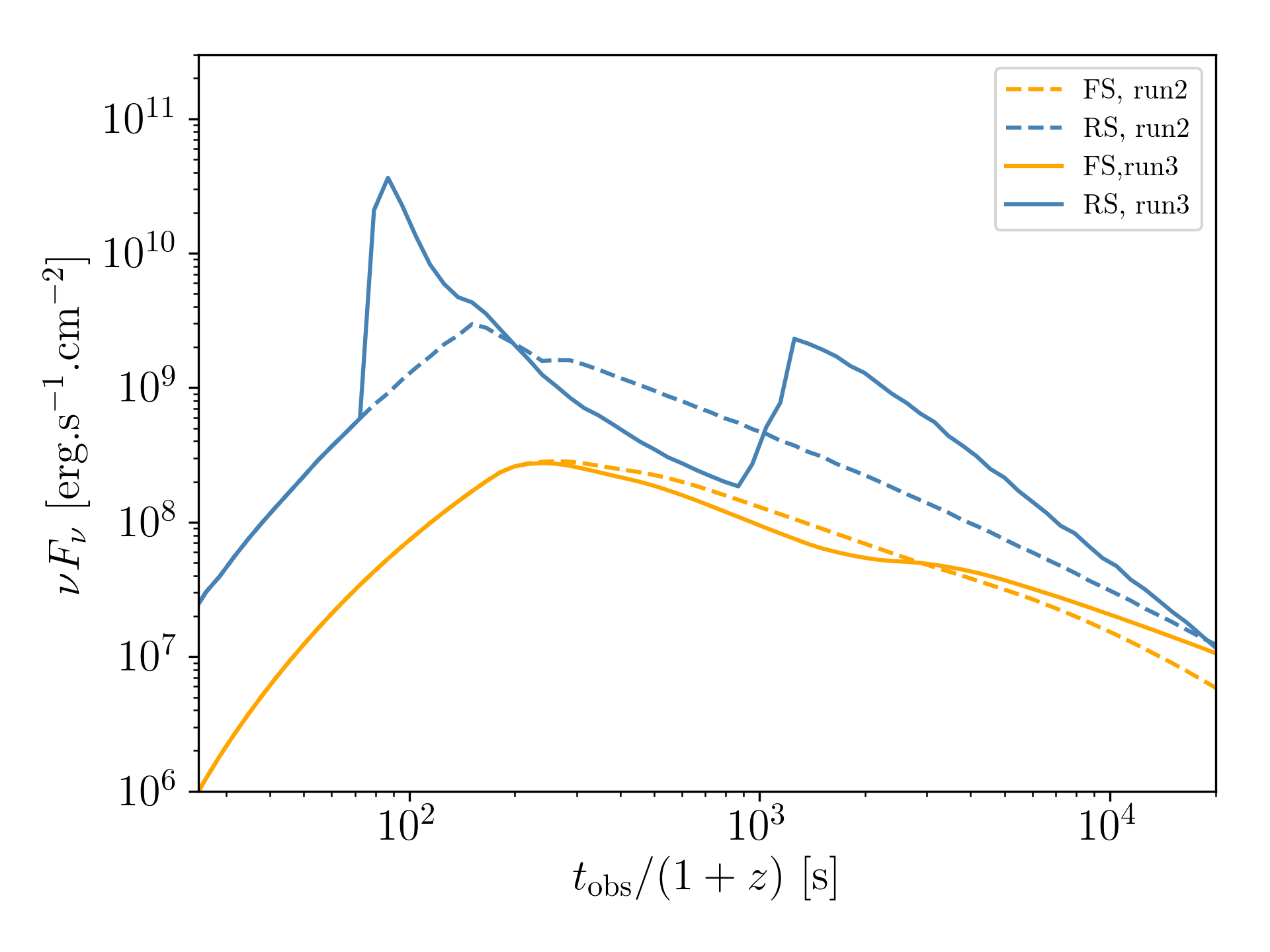}
      \caption{X-ray light-curves for the FS (orange) and RS (blue) at 1 keV for run2 (gradual central engine shutdown) and run3 (chaotic central engine shutdown) for a non-jetted outflow. The RS displays flaring behavior in run3 at $t_\mathrm{obs} \sim 3 \times 10^3$ s. The FS shows a rebrightening at similar times, due to refueling by reflected forward shocks. The spike in the RS at $t_\mathrm{obs} \sim 160$s is associated with the internal shocks and akin to prompt emission mechanisms.}
      \label{fig:LD_lc}
    \end{figure}

\begin{figure}
      \centering
      \includegraphics[width=0.49\textwidth]{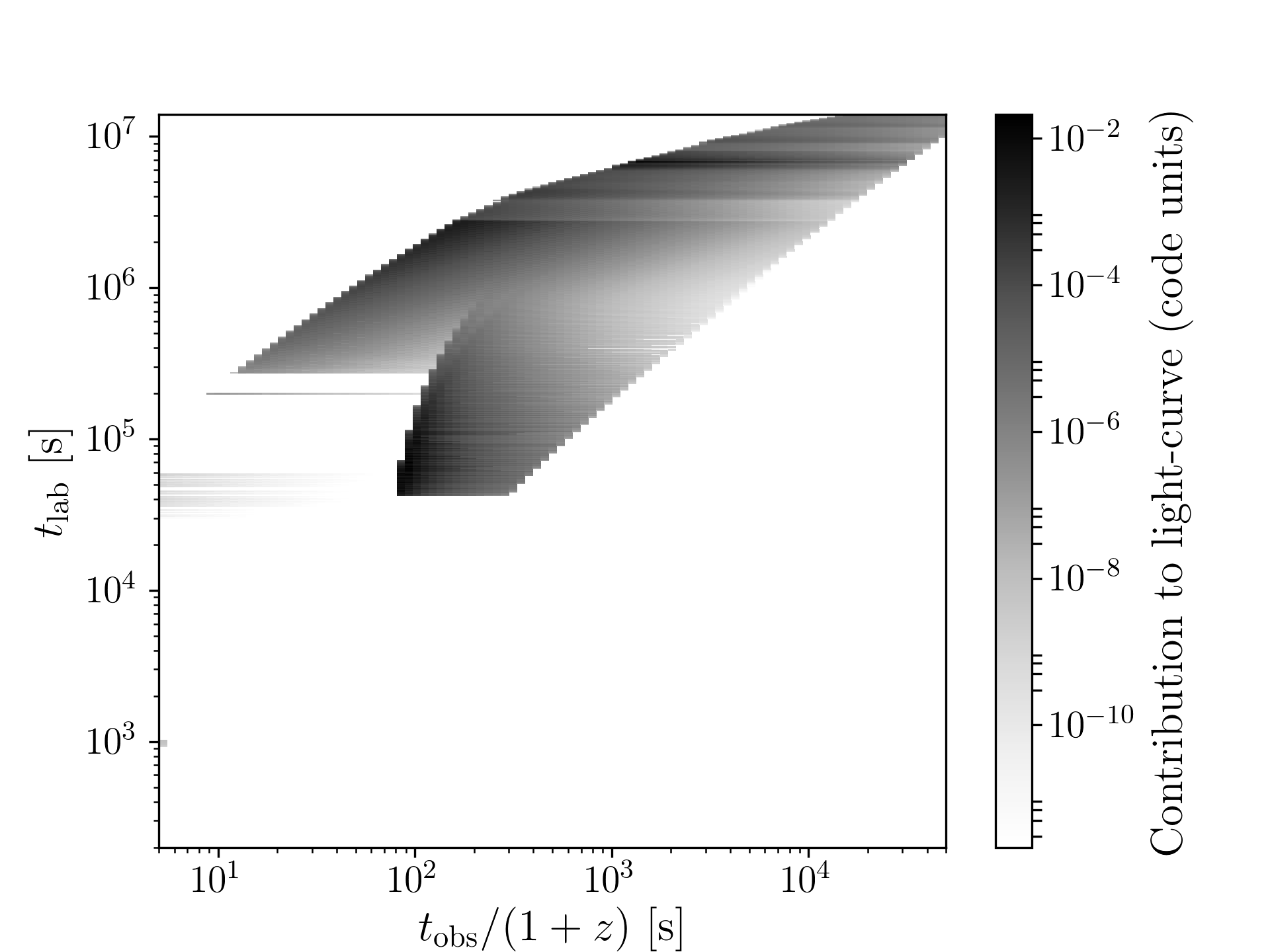}
      \caption{Normalised contribution of every data dump to the light curve for run3. The light-curve is simply the stack of all these contributions.}
      \label{fig:light_map_spherical}
    \end{figure}

    In figure \ref{fig:LD_lc} we show the X-ray light-curves at 1 keV obtained in the case of a spherical outflow and isotropic local emission for run2 (tail) and run3 (non-monotonic). These are in very good agreement with the results from \citetalias{Lamberts2017a}. The emission is computed with the same value of $\epsilon_e = 0.1$ in the FS and RS,  but $\epsilon_B = 0.1$ in the RS and $\epsilon_B = 10^{-6}$ in the radiatively inefficient FS. We discuss in section \ref{sub:emission_from_the_shocked_external_medium} the validity of the choice of values for the FS. 
    The lower emission at very early times compared to \citetalias{Lamberts2017a} before 100~s in our simulations is due to our decision not to include the variable "head" region that they use in their simulations, as it is not linked to the flaring phenomenon. We obtain very similar results displaying a first spike at $\sim 160$ seconds corresponding to emission from the internal shocks as the dense shell is being formed (we do not include prompt emission or a steep decay phase at this time in our synthetic light curves. The visibility of this first spike would directly depend on the characteristics of those features). Indeed, fig. \ref{fig:light_map_spherical} reports the contribution of all lab times to the final light-curve and shows that all the radiation in this first spike is emitted mostly between $t_\mathrm{lab} \sim 4 \times 10^{4}$ s and $t_\mathrm{lab} \sim 5 \times 10^{5}$ s, the time of the formation of the dense shell in our dynamical simulations (recall fig. \ref{fig:run3_dynamics}). The arrival time of this spike is comparable to the time at which the corresponding emitting material was ejected by the central engine in the lab frame, in the same fashion as the internal shock mechanism invoked for burst emission \citep{Rees2000,Zhang2002}. A second bump is visible in the light curve at $t_\mathrm{obs} \sim 3 \times 10^{3}$ s and corresponds to increased emission from the RS as it traverses the dense shell. Again, this is highlighted by the contributions map showing that this radiation is now only produced at very late lab times corresponding to the RS crossing the dense region. This emission will be able to peak above a radiatively inefficient FS in a flare-like manner. Allowing the microphysics to differ in the ejecta and the external medium thus makes it possible for flares to have a RS origin. The FS emission eventually displays a rebrightening occurring just after the flare. This rebrightening is expected as a consequence of the reflected FS formed by the interaction of the RS with the dense layer and is stronger in our simulations than in that of \citetalias{Lamberts2017a} because of the second forward reflected shock.
    However, we will see in section \ref{sub:short_flares_in_a_structured_jet} that the rebrightening actually completely disappears when taking into account the constraints on flare timescale in jet setups with an angular patch, side-stepping the potential shortcoming of this scenario that would have associated flares with rebrightenings in GRB afterglows.

    \begin{figure}
      \centering
      \includegraphics[width=0.49\textwidth]{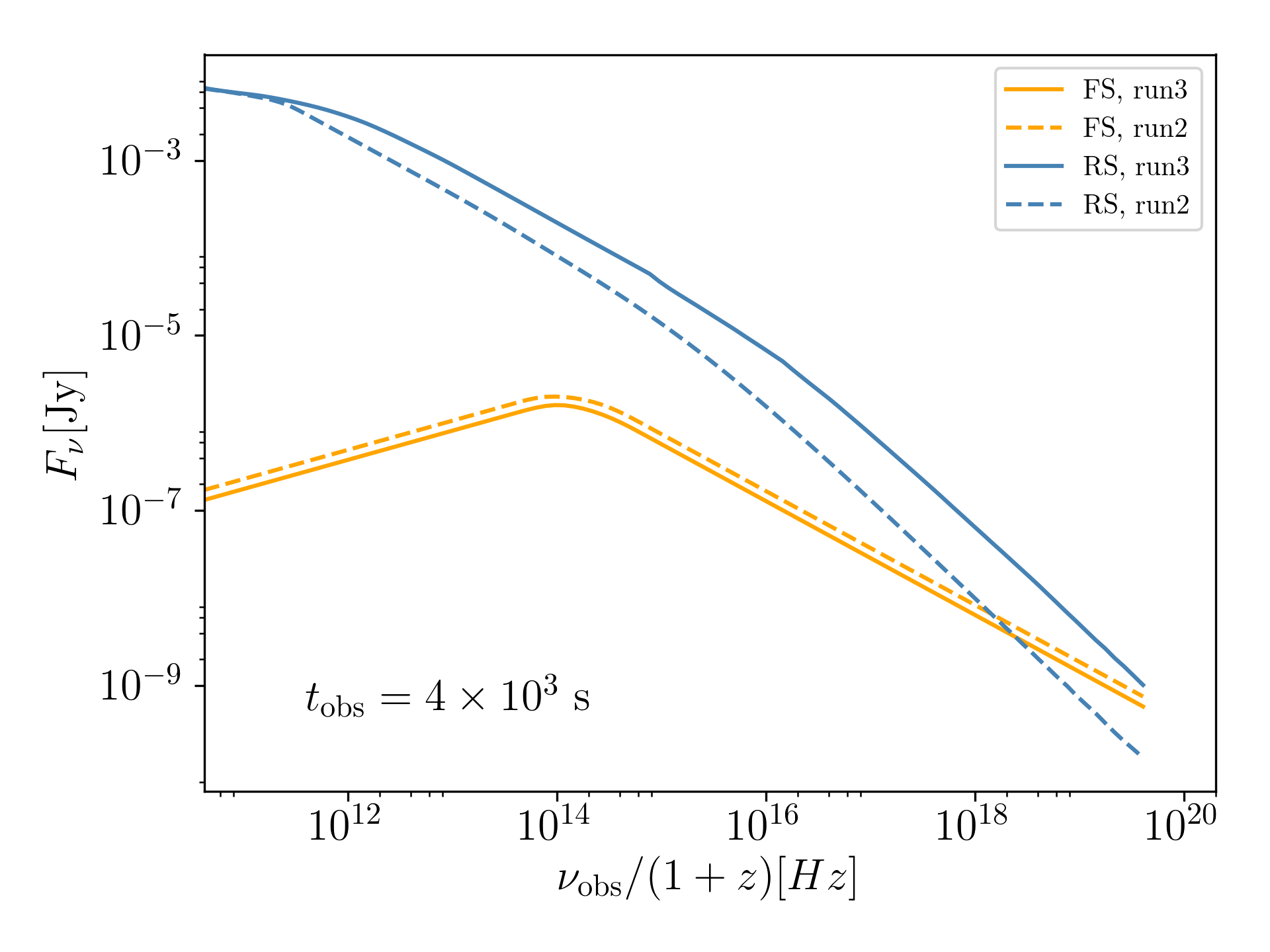}
      \caption{Spectral energy distribution for run2 and run3 at the time of the flare $t_\mathrm{obs} = 4 \times 10^3$ s. In this figure, $\epsilon_{B,\mathrm{RS}} = 0.1$ and $\epsilon_{B,\mathrm{FS}} = 10^{-6}$.}
      \label{fig:spectra_spherical}
    \end{figure}

    Thanks to our emission calculation prescription, we are able to produce accurate synthetic broadband SED for the afterglow emission. Figure \ref{fig:spectra_spherical} shows the FS and RS spectra for run2 (tail) and run3 (non-monotonic). As expected for early times, we get a harder spectrum for the FS than for the RS, with a dominating RS at low frequencies. From these SEDs, we can see that the emission increases for all frequencies in the RS at the time of the flare. However, the cooling break also increases from $\nu_{c,\mathrm{run2}} \sim 6 \times 10^{15}$ Hz to $\nu_{c,\mathrm{run3}} \sim 3 \times 10^{16}$ Hz. As a result, the relative flux change in optical is smaller than the change in X-ray, as optical frequencies are placed below the cooling break, and X-ray above. Depending on how radiatively efficient the FS is, it could allow the flare to dominate in X-ray, but not in optical, as suggested by observations. Finally, it appears that the rebrightening FS in X-ray is linked to a second radiating component with a very high cooling frequency. However, as mentioned before, we will show later that this emission can be totally suppressed if the dynamical perturbation is constrained to a small opening angle.


  \subsection{Short flares from a localised angular emission region} 
  \label{sub:short_flares_in_a_structured_jet}

    As mentioned in \citetalias{Lamberts2017a} and shown in the previous section, the timescale of this flare is longer than observed, with $\Delta t_\mathrm{flare} / t_\mathrm{flare} \sim 1$ instead of the $\Delta t_\mathrm{flare} / t_\mathrm{flare} \sim 0.1 - 0.3$ reported by \citet{Chincarini2010}. \citetalias{Lamberts2017a} choose to invoke anisotropy of the emission in the co-moving frame in order to reconcile simulations with observations, essentially limiting the emission to a smaller solid angle. However, this constraint could very well be produced by the geometry of the jet itself. We present here a different approach that avoids having to introduce additional constraints on the microphysics of emission. Here, the short timescale of flares translates into constraints on the spatial extent of the emitting region. In the late activity of the central engine scenario, this constraint is satisfied as the emission is taking place at very small radii close to the remnant, such that the timescale of the flare is directly proportional to the variability timescale of the activity of the central engine. As a result, it is possible to reach the small values of $\Delta t_\mathrm{flare} / t_\mathrm{flare}$ observed rather easily, but it is more difficult to justify the ad hoc requirement of an evolving timescale for the central engine activity \citep{Kobayashi2005}.
    In the reverse shock scenario, the criterion on spatial extent can be fulfilled in the angular direction. The variability of the initial ejection can leave an imprint on the radial as well as the angular velocity profile of the ejecta. In this situation, an over-dense region of material could form only in a "filament" of ejecta in a small angular region, with which the RS would eventually interact.

\begin{figure}
      \centering
      \includegraphics[width=0.3\textwidth]{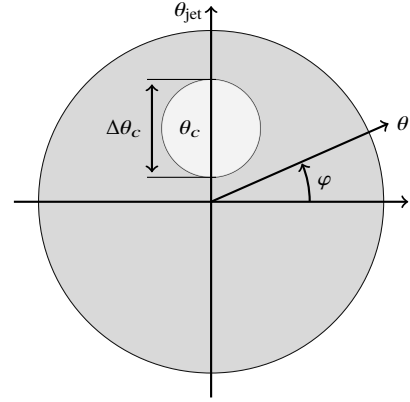}
      \caption{Schematic representation of the angular setup for the jet viewed on-axis. The shaded region represents the non-perturbed dynamics, whereas the white region corresponds to the perturbed "filament" responsible for the flare.}
      \label{fig:theta_ini}
    \end{figure}

    We investigate this possibility by limiting the perturbed dynamical setup to a small angular region at the initial state. The angular setup is described in figure \ref{fig:theta_ini}. We expect this setup to allow the cooling timescale to dominate the flare decay time, suppressing the curvature effect. We show that the reverse shock scenario can produce flares in filaments away from the line of sight as well as on the line of sight, before focusing on flaring behavior aligned with the observer.

    In practice, we use the dynamics results of run2 and run3 and assign them to different angles in the light-curve and spectral reconstruction, to form a non-perturbed (NP) and a perturbed (P) region, respectively. Later in this work, we will assign different dynamics outputs to the P region and will refer to the integrated light-curve and spectra over P and NP by the name of that dynamical output (since NP is unchanged and remains run2). In the on-axis ($\theta_c = 0$) case, the dynamics $\mathcal{D}(\theta)$ eventually come down to:
    \begin{align}
      \mathcal{D}(\theta) = 
      \begin{cases}
        \mathcal{D}_\mathrm{P}, & 0 \leq \theta < \Delta \theta_c \\
        \mathcal{D}_\mathrm{NP}, & \Delta \theta_c \leq \theta < \theta_\mathrm{jet}
      \end{cases},
    \end{align}
    with $\Delta \theta_c$ the opening angle of the P region and $\theta_\mathrm{jet}$ the jet opening angle. As a first approach, and provided that $\Delta \theta_c$ is sufficiently large, we can thus limit ourselves to post-processing of 1D dynamical simulations as we expect the P region to be non-causal at least until the reverse shock has crossed the dense shell. The NP region that constitutes the bulk of the jet is also non-causal as we model its evolution pre-jet break. From these considerations on the validity of our setup, we extract the following constraints:
    \begin{align}
      & \Delta \theta_c \gtrsim 0.01 \text{ rad},\\
      & \theta_\mathrm{jet} \lesssim 0.1 \text{ rad}.
    \end{align}

\begin{figure}
      \centering
      \includegraphics[width=0.49\textwidth]{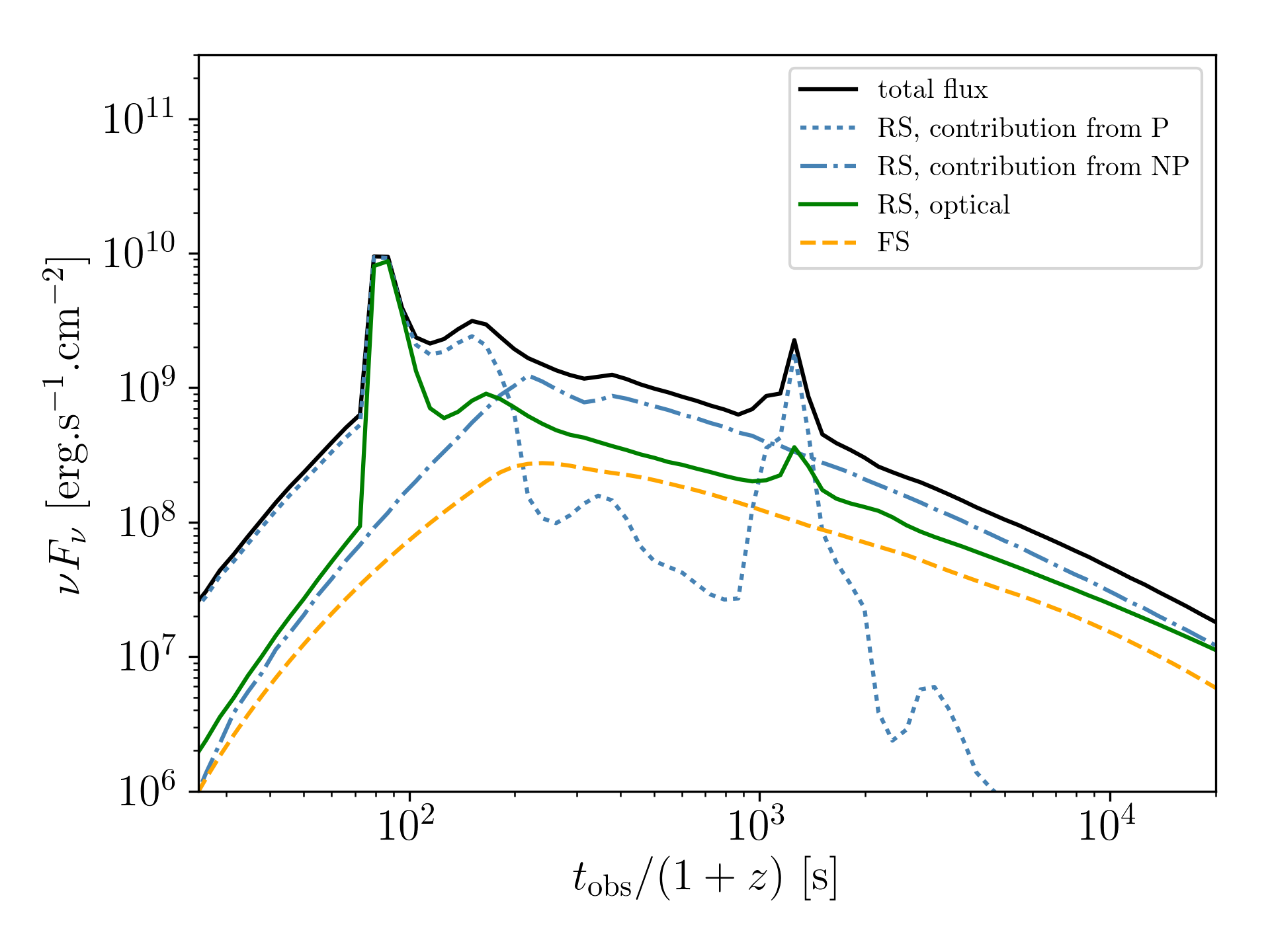}
      \caption{X-ray light-curves at 1keV and optical light-curve (green) for a jet perturbed in an angular patch ($\theta_\mathrm{jet} = 0.1$ rad, $\theta_c = 0$, $\Delta \theta_c = 0.015$ rad, black solid curve) and for a non-perturbed jet (black dashed curve). The blue dashed and dot-dashed curves are the contributions to the RS from the NP and P regions respectively. The solid green curve is the optical light-curved for the flaring setup. The timescale of the flare is significantly reduced in comparison with globally perturbed setups.}
      \label{fig:struc_setup_lc}
    \end{figure}

    In the results presented in fig. \ref{fig:struc_setup_lc}, we chose $\Delta \theta_c = 0.015$ and $\theta_\mathrm{jet} = 0.1$. The value of $\theta_\mathrm{jet}$ corresponds to a canonical observed jet opening angle \citep{Ryan2015}. As expected, we observe a decrease of the flare decay time, in better accordance with observations. 
    We can then empirically derive a final constraint on $\Delta \theta_c$ corresponding to the maximum value it can take in order to retrieve the flare timescales measured from observations. From our simulations, we record that $\Delta t_\mathrm{flare} / t_\mathrm{flare}$ starts increasing for $\Delta \theta_c \gtrsim 0.01 \mathrm{~rad}$, which actually corresponds to the maximum expected size for the patches allowed by transverse causal contact $1/\Gamma$. This limit supports the validity of our scenario since flares with longer timescales than that observed can't actually be produced.

\begin{figure}
      \centering
      \includegraphics[width=0.49\textwidth]{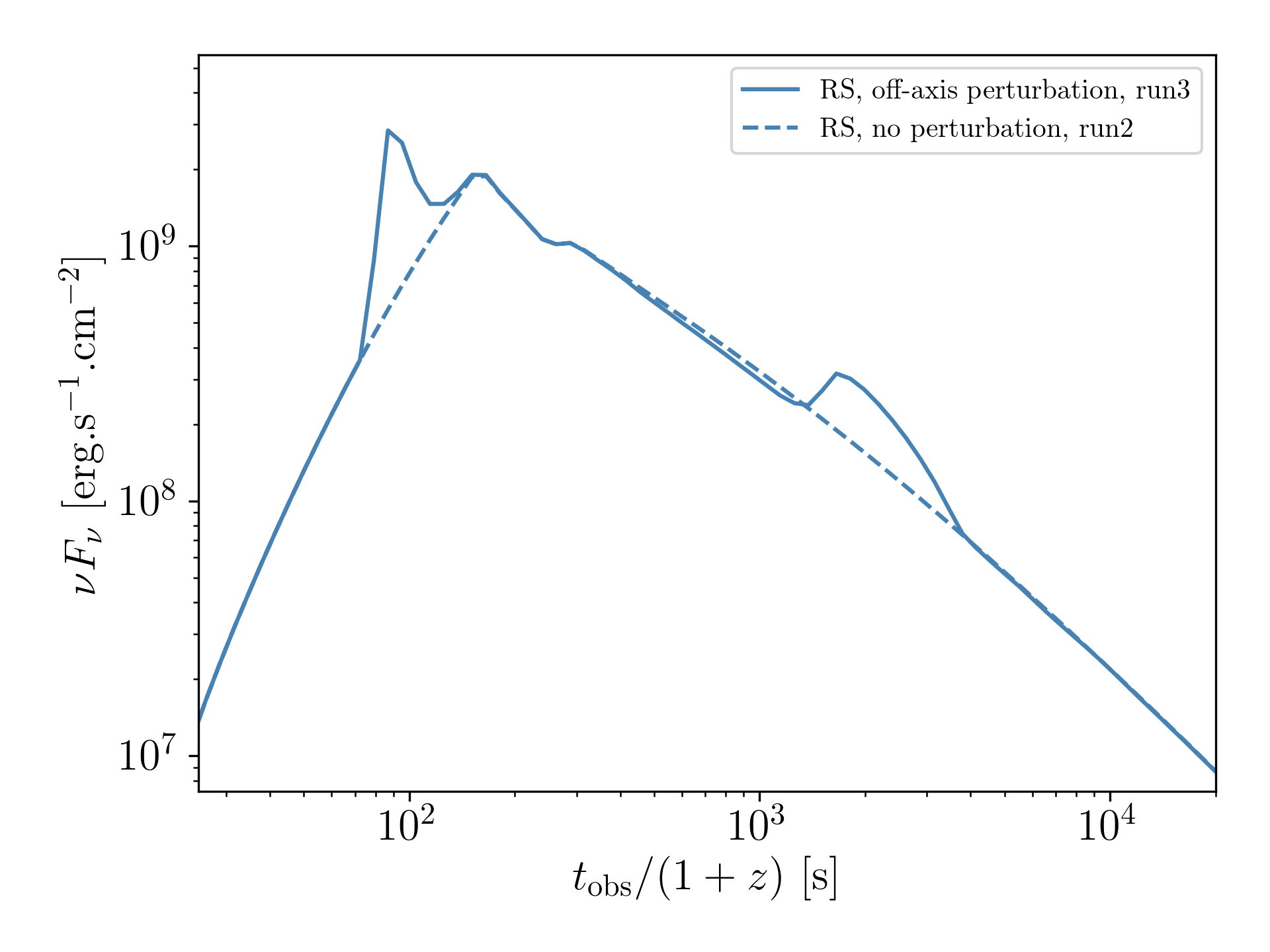}
      \caption{X-ray light-curves at 1keV for a jet with a perturbed angular patch away from the line of sight ($\theta_\mathrm{jet} = 0.1$ rad, $\theta_c = 0.02$, $\Delta \theta_c = 0.03$ rad, blue solid curve).}
      \label{fig:off_axis_flare}
    \end{figure}

    Fig. \ref{fig:off_axis_flare} also shows that this setup can produce X-ray flares with perturbations away from the line of sight, as visible when offsetting the P region by an angle $\theta_c = 0.02$. Due to curvature effects, the timescale of the flare increases when moved to higher angles. However, the flux of the the flare also decreases in that case due to simple Doppler beaming effects. As the flare coming from the RS needs to peak above the FS emission, this latter effect is also responsible for a decrease in timescale. It also means that perturbations away from the line of sight will sometimes not give rise to flares at all, causing concerns about the timescales of flaring behavior to be irrelevant in these directions. Additionally, this effect can be responsible for the large scatter in $\Delta F / F$ measured in observations \citep{Chincarini2010,Margutti2011}.

\begin{figure}
      \centering
      \includegraphics[width=0.49\textwidth]{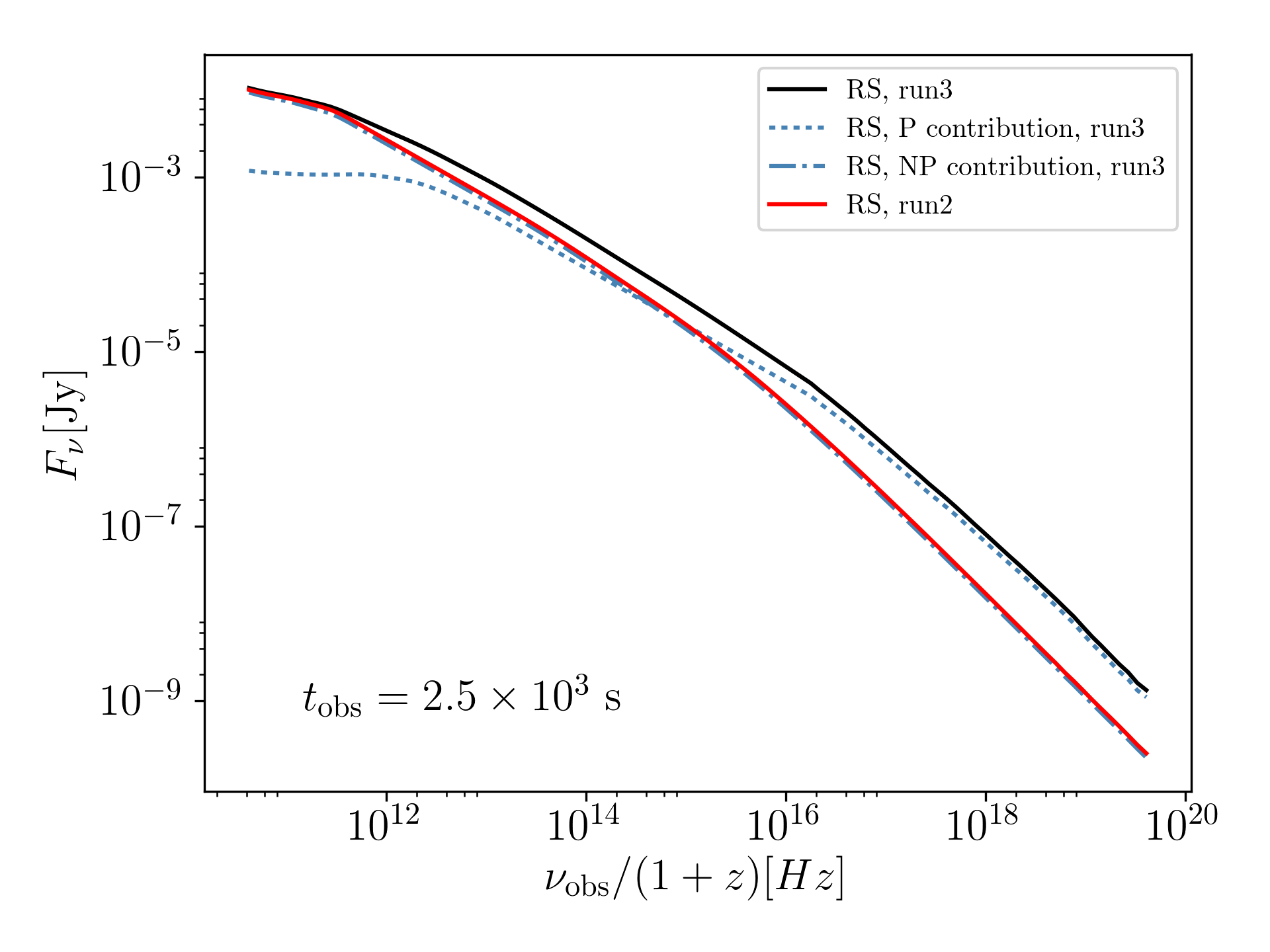}
      \caption{Simulated RS spectra at the flaring observer time $t_\mathrm{obs} = 2500$s for the flaring (solid black line) and non-flaring (solid red line). The dotted and dash-dotted blue lines show the contributions of the P and NP regions to the total RS spectrum in the flaring setup.}
      \label{fig:spectra}
    \end{figure}

    Figure \ref{fig:spectra} shows the contributions of the P and NP regions to the spectral energy distribution for the case aligned with the observer at the time of the flare. Just like in the spherical blast-wave case (fig.~\ref{fig:spectra_spherical}), the X-ray flare is due to an increase in both the peak frequency and the cooling frequency in the radiation from the P component. The significant increase in the cooling frequency in the P component is responsible for a stronger flare signature above the cooling break. This is responsible for the flare being much smaller in optical than in X-ray.

  \subsection{Evolution and influence of the shocked external medium} 
  \label{sub:emission_from_the_shocked_external_medium}

    We presented here the first set of simulations with a realistic treatment of the external medium density for $n_0 = 1\text{~cm}^{-3}$. Even though the RS is weaker than for larger values of $n_0$, these simulations first show that the setup still produces flares for a more diffuse external environment and is not dependent on the very high density of previous simulations. Moreover, the lower density causes the reverse shock to form later in the lab frame, causing the flares to appear later in the observer frame. This is a first encouraging result towards the ability of this setup to produce late flares. We study late flares in more detail in section \ref{sub:late_flares}

    Flares can peak over the FS thanks to the difference in radiative efficiency between the ejecta and the external medium. In our simulations, we chose $\epsilon_{B,\mathrm{RS}} = 0.1$ and $\epsilon_{B,\mathrm{FS}} = 10^{-6}$. Even though a change in the value of $\epsilon_e$ would yield the same result, the constraints are more stringent on this parameter with $\epsilon_e \sim 0.13 - 0.15$ \citep{Beniamini2015}. On the contrary, \citet{Santana2014} measure a wide scatter of $\epsilon_B \sim 10^{-8} - 10^{-3}$ in the FS, with a median for $\epsilon_B \sim 10^{-5}$. This coincides with our choice of value of $\epsilon_B$.

    As mentioned in the previous section, the interaction of the RS with the dense region in the tail of the ejecta is responsible for a rebrightening in the FS at the observer time of the flare. 
    As it turns out, and is shown in fig. \ref{fig:struc_setup_lc}, the rebrightening is strongly suppressed when constraining the perturbation to a small angular region, for various reasons. First, the increased emission now comes from a smaller region, causing the total flux to be lower. Second, the region being narrow, the rebrightening is only visible while the beaming angle is smaller than the perturbed region opening angle. When this is not the case anymore, a break in the contribution of the perturbed region to the light-curve causes its emission to drop below the underlying non-perturbed jet emission.

    The fact that the rebrightening disappears in the case of a narrow emitting region supports the validity of our setup, as it would otherwise imply that all flares have to be associated with a subsequent rebrightening.


\subsection{Late flares}
\label{sub:late_flares}

The currently accepted paradigm for X-ray flares in GRB light curves invokes late activity of the central engine, either from late-time fallback accretion, or from energy injection from a NS remnant. Various works have shown the ability of this model to reproduce the observed flare properties. However, a significant number of GRBs display flares at times as late as $10^{5}$s after the burst \citep{Burrows2005}. The scenario involving late activity of the central engine runs into a limitation there as the timescale on which the flare shows in observer time is equal to that over which the central engine reactivates. This puts severe constraints on the nature and stability of the remnant. \citet{Piro2019} show a rebrightening around 160 days at $3.2\sigma$ in the X-rays relative to the underlying broadband continuum for the famous gravitational counterpart afterglow of GRB170817A (the significance decreases to $2.8\sigma$ if the X-rays are analysed in isolation \citep{Piro2019,Hajela2019}).
They suggest energy injection from a hypermassive NS, which further emphasizes the need for an explanation for flares at very late time that does not rely on assumptions about the nature of the remnant. In this section, we show how a LLRS can be responsible for flares at early and late times. 

\begin{figure}
  \centering
  \includegraphics[width=0.49\textwidth]{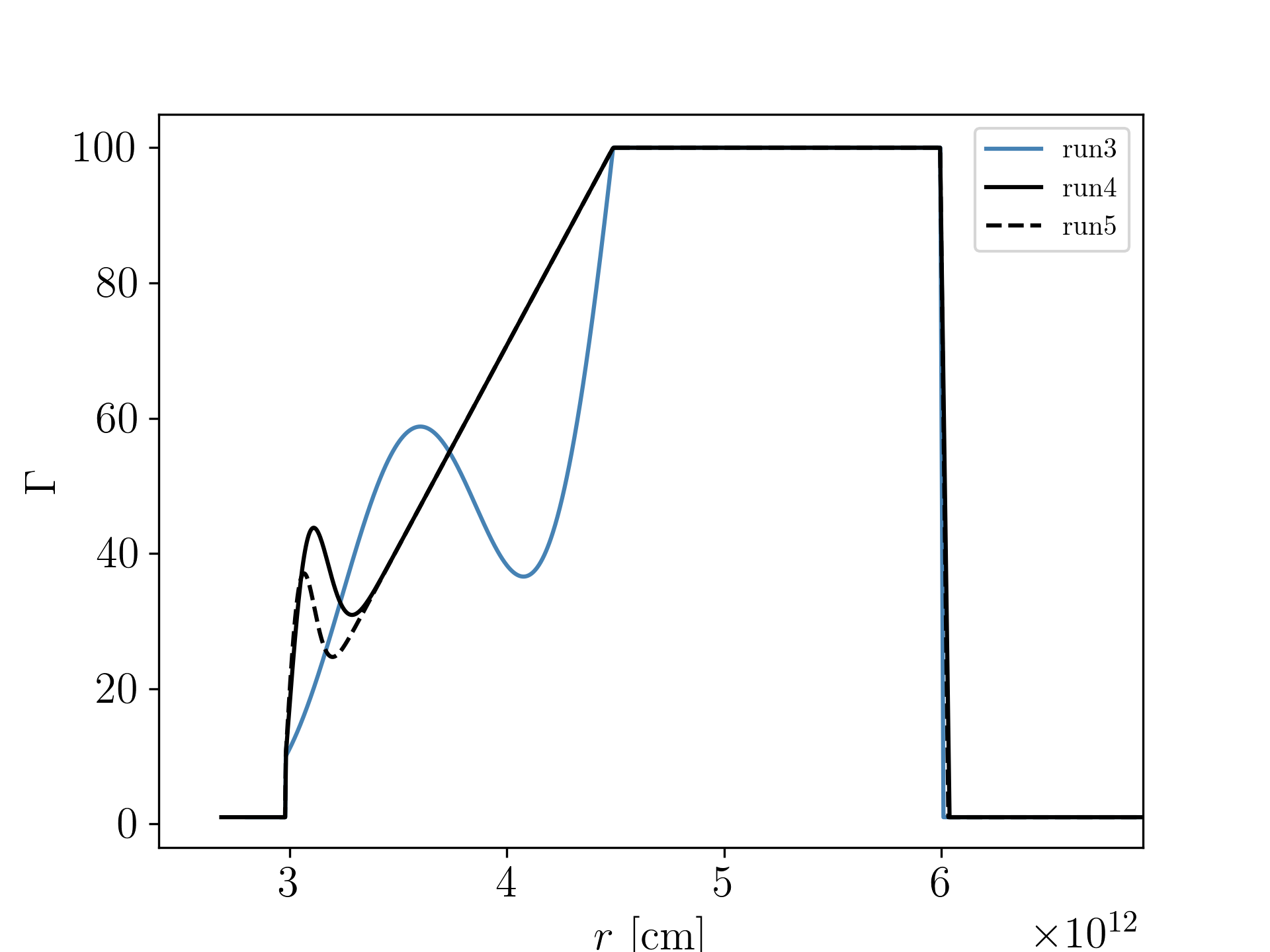}
  \caption{Initial Lorentz factor radial profile for run4 and run5. The gaussian bump is moved further at towards the back of the ejecta in comparison to the oscillation visible in run3.}
  \label{fig:gauss_init_profile}
\end{figure}

We investigate how our setup can reproduce flares at later times by modifying the previous dynamical setup as follows: The radial Lorentz factor profile still displays a flat head region, followed by a smoothly decreasing region corresponding to gradual shutdown of the central engine. The chaotic shutdown of the engine is now modeled by a Gaussian profile added to the gradual shutdown. This is done by replacing $f(x)$ in eq. \ref{eq:initial_gamma_profile} by $g(x)$ such that:
\begin{align}
  g(x) = A e^{\left(\frac{x - x_0}{\sigma_0}\right)^{2}}.
\end{align}
This bump can be moved left or right to model earlier or later ejection times by changing the value of $x_0 \in [0,1]$, with $\sigma_0$ controlling the width of the bump.
The Lorentz factor initial radial profiles are shown in fig.~\ref{fig:gauss_init_profile} and the corresponding parameters are reported in table \ref{tab:late_run_parameters}. $A$ and $\sigma_0$ are kept the same in both run so as to investigate whether $x_0$ would have an influence on its own on $\Delta F / F$ for the produced X-ray flare. The strength of this approach lies in that the shell still has the same width of 100 light-seconds. The late variability in the afterglow can then be imprinted onto the dynamics at the time of the ejection and does not rely on late activity of the central engine.

\begin{table}
  \small
  \begin{tabular}{lrrrl}
    Run & $ A $ & $x_0$ & $\sigma_0$ & Description \\
    \hline
    run4 & 8 & 0.01  & 0.005 & early bump\\
    run5 & 8 & 0.001 & 0.005 & late bump\\
    \hline
  \end{tabular}
  \caption{Run Parameters for late flare setups.}
  \label{tab:late_run_parameters}
\end{table}

\begin{figure}
  \centering
  \includegraphics[width=0.5\textwidth]{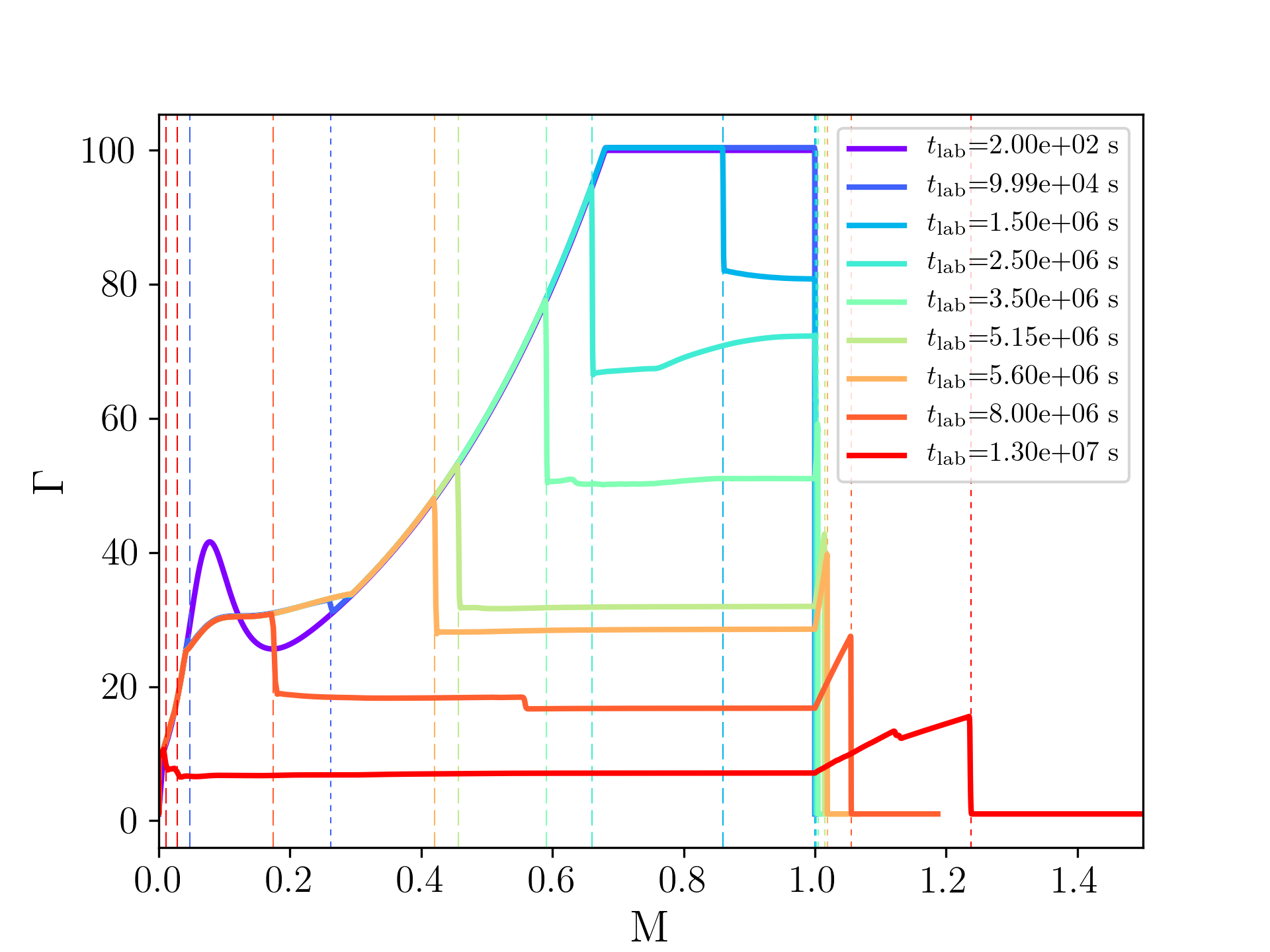}
  \includegraphics[width=0.5\textwidth]{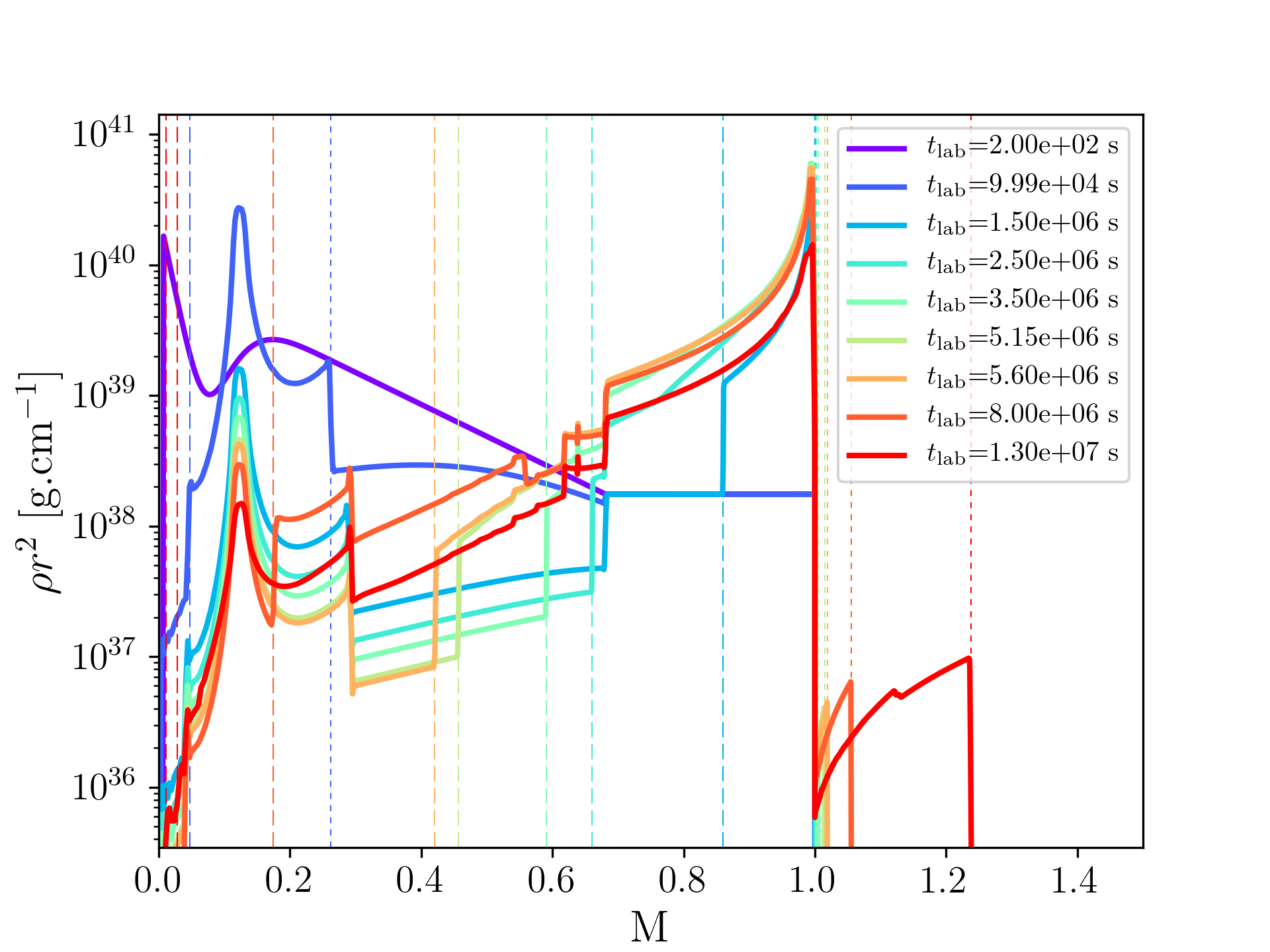}
  \caption{Lorentz factor (top) and density (bottom) as a function of cumulative mass for run4 (chaotic shut down of the central engine leading to a region of negative velocity gradient in the trail of the ejecta. Dashed vertical lines show the positions of reverse shocks and dotted vertical lines the positions of forward shocks.)}
  \label{fig:run4_dynamics}
\end{figure}

\begin{figure}
  \centering
  \includegraphics[width=0.5\textwidth]{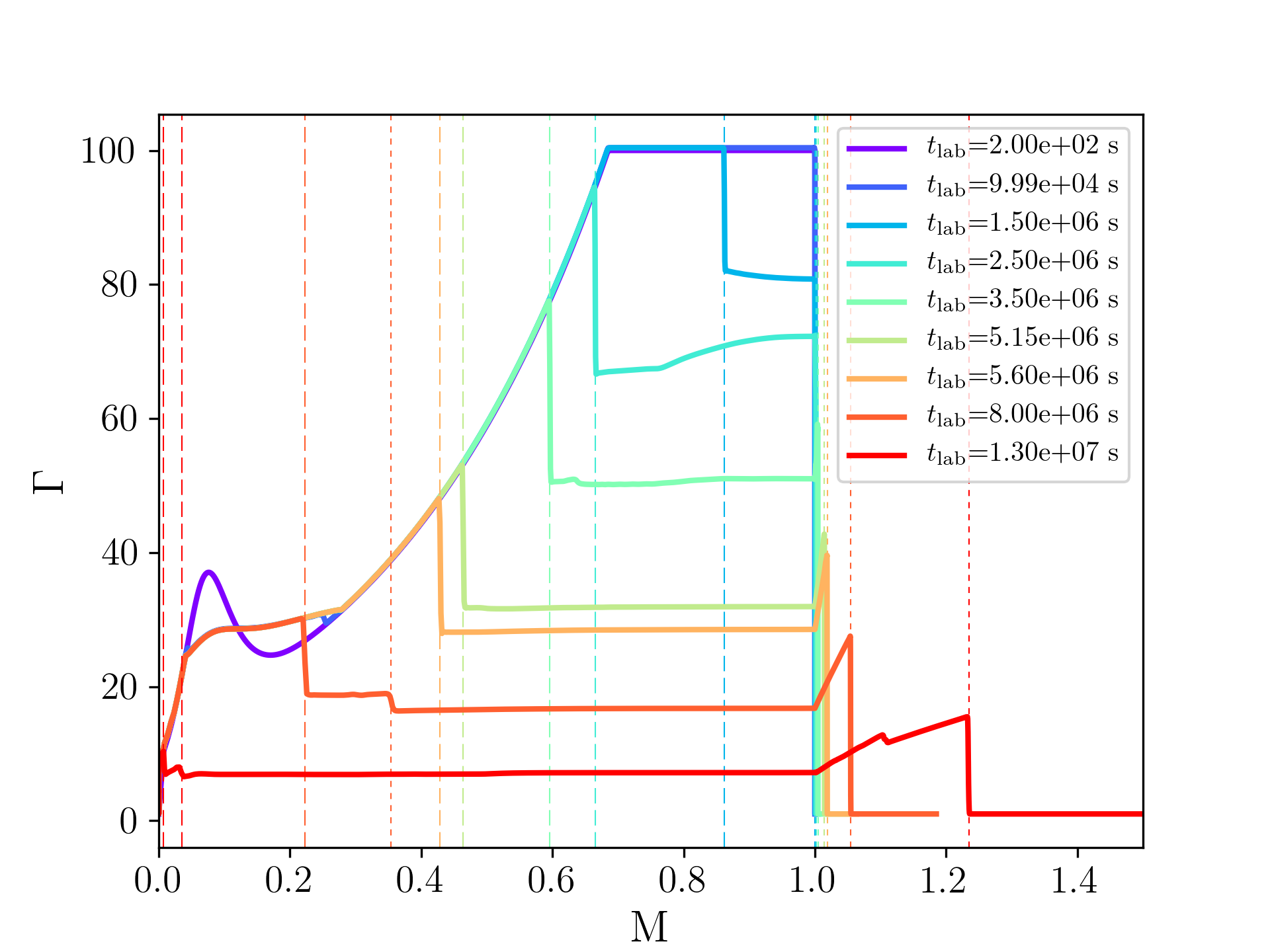}
  \includegraphics[width=0.5\textwidth]{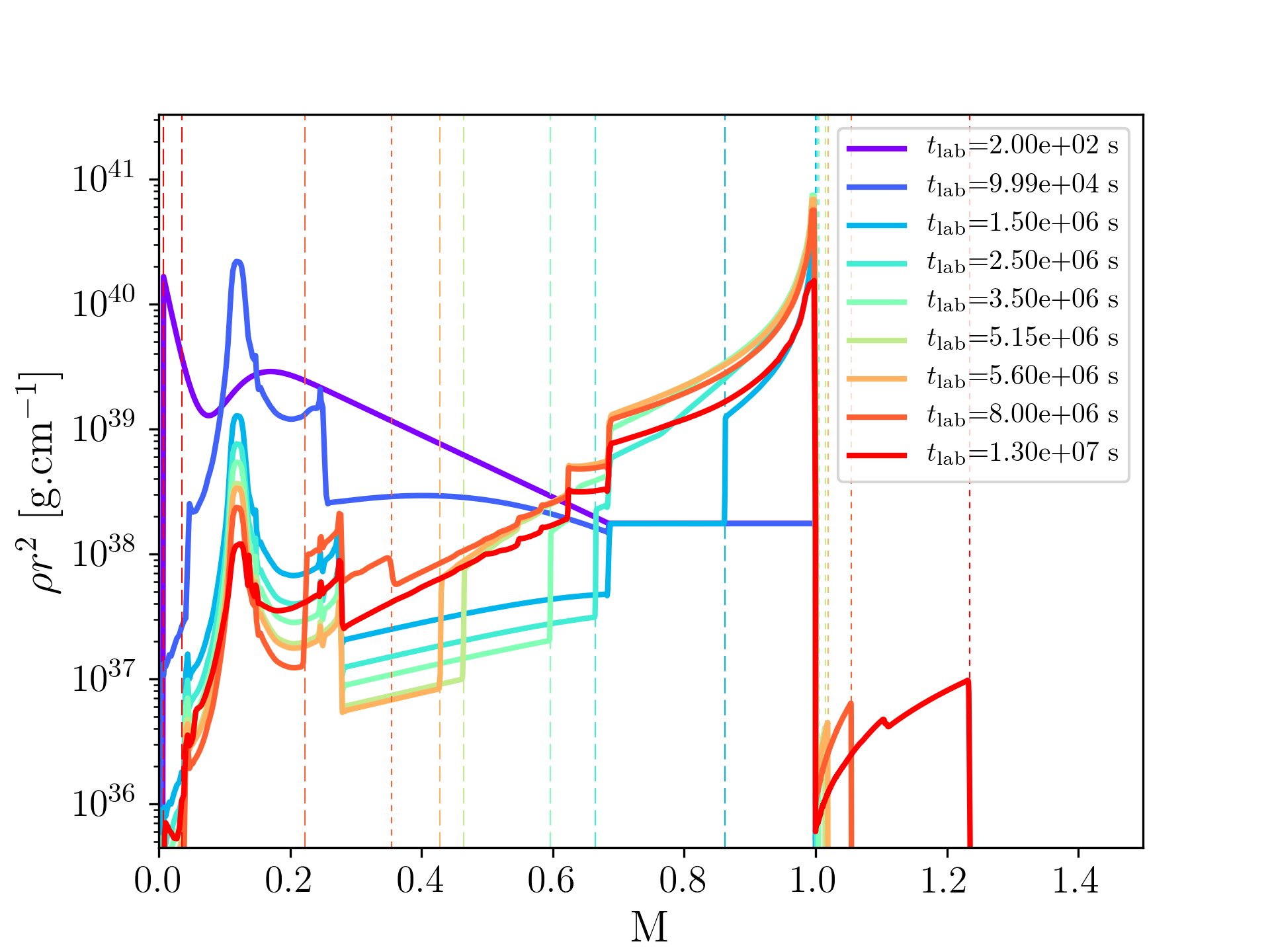}
  \caption{Lorentz factor (top) and density (bottom) as a function of cumulative mass for run5 (chaotic shut down of the central engine leading to a region of negative velocity gradient in the trail of the ejecta. Dashed vertical lines show the positions of reverse shocks and dotted vertical lines the positions of forward shocks.)}
  \label{fig:run5_dynamics}
\end{figure}

\begin{figure}
  \centering
  \includegraphics[width=0.49\textwidth]{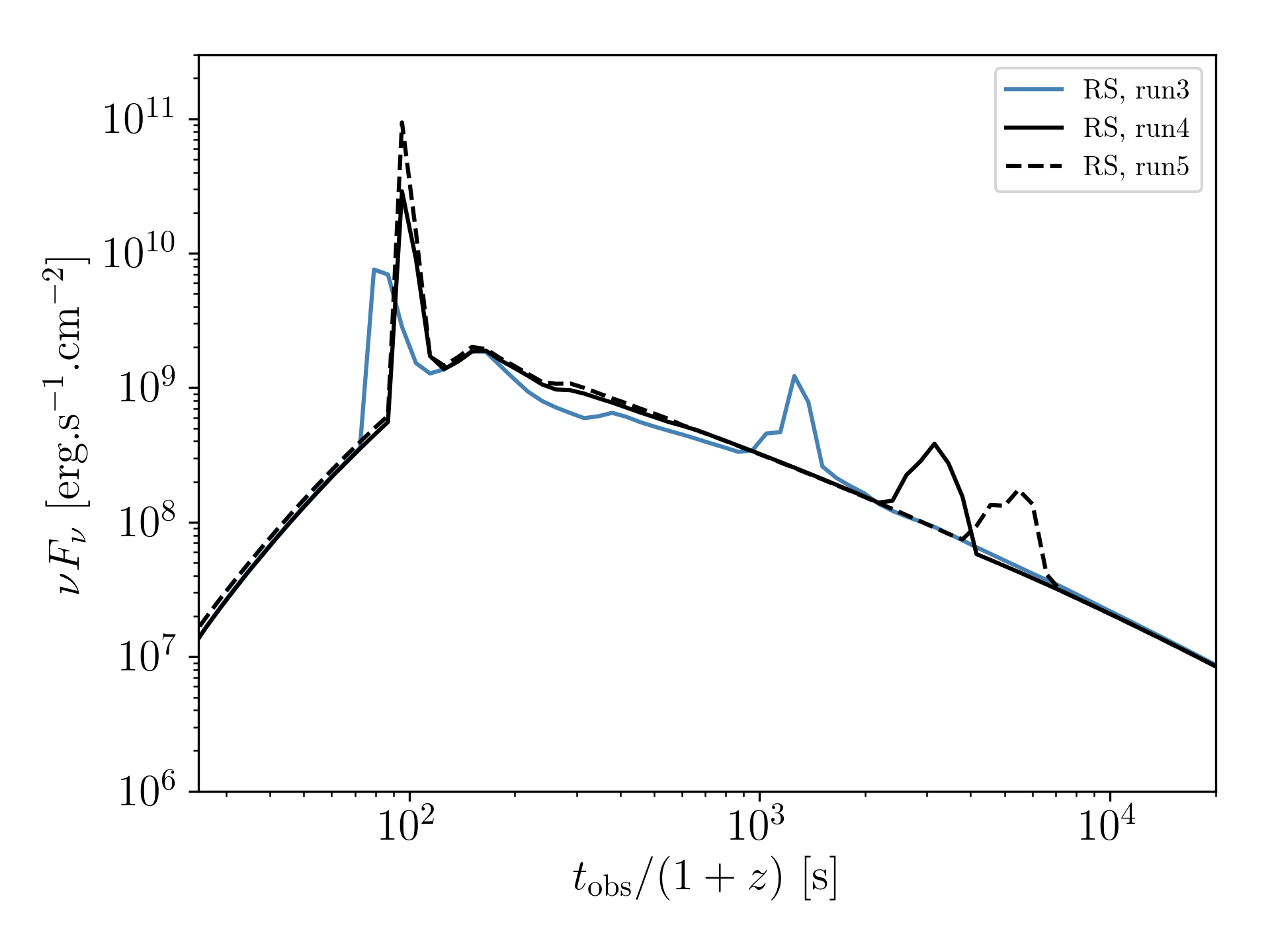}
  \caption{X-ray light-curves for run3, run4, run5. The flare moves to later observer times as non-monotonic region of velocity is moved further at the back of the ejecta. (Preliminary figure)} 
  \label{fig:flares_gauss}
\end{figure}

The dynamics are very similar to the dynamics described in section \ref{sec:hydrodynamical_evolution} and are shown in figures \ref{fig:run4_dynamics} and \ref{fig:run5_dynamics}. As in previous simulations, the region of negative velocity gradient in the tail leads to the formation of a dense region of ejecta surrounded by two internal shocks. Again, the RS emission is powered up as it traverses this shell and is responsible for a flare. Fig.~\ref{fig:flares_gauss} shows the synthetic X-ray light-curves for this setup for various values of $x_0$. As expected smaller values of $x_0$ lead to later peak times of the produced flares. The significant result of these simulations is that the timescale ratio of the flares remains the same, independent of the value of $x_0$, showing that the cooling timescale remains the same regardless of the arrival time of the flare, making the reverse-shock scenario for flares all the more robust. Another striking feature is that the flux variability $\Delta F / F$ is the same for run4 and run5, for flares occurring at different times, in accordance with observations. This suggests that it is possible to produce flares up to an arbitrarily long time after the burst if one considers a gradual shutdown of the central engine all the way to Lorentz factors of order unity, over a timescale consistent with a short-lived central engine, while still fulfilling the flux variability and flare timescale requirements.

\section{Discussion}

We demonstrated the ability of the RS to produce flares at both early and late times as it interacts with a dense region in the ejecta created by the non-monotonic initial radial velocity profile. These flares can arise both in a direction aligned and non-aligned with the line of sight and satisfy the timescale constraints given by observations provided that the region producing them is narrow enough. However, some aspects remain to be investigated.

\subsection{Limitations to dynamical simulations} 
\label{sub:limitation_to_dynamical_simulations}

Even though we offer significant improvement over previous studies thanks to the moving mesh approach, both in terms of efficiency and accuracy, because we are able to run more simulations with a wider range of more realistic parameters, our description of the dynamics still relies on a number of simplifying hypotheses. First, we are running 1D-simulations that ignore all lateral motion. This is a good approximation at early times as the jet is non-causal but becomes less valid as the blast-wave decelerates at later times, leading to spreading and growing shear-driven lateral instabilities. However, since we are mostly interested in the dynamics of the reverse shock at Lorentz factors of a few tens, 1D simulations remain a good approximation all the way to crossing time. Still, given the complexity of the dynamics involved, and  the need to introduce angular structure to explain the observed timescale for flares, we expect that this study will benefit from comparison with 2D simulations in the future.

Second, we neglect in these simulations the dynamical effect of potential magnetic fields and assume a low magnetisation of the ejecta. At late stages following launching, magnetic fields are not expected to be dynamically dominant \citep[see e.g.][for a recent review]{Granot2015}. We consider residual magnetic fields that are small enough such that they do not influence the strength of the reverse shock, as a strong magnetisation would hinder the propagation of the RS, making this scenario unviable \citep{Giannios2008,Mimica2009}. 


\subsection{Validity of the flaring setup} 
\label{sub:validity_of_the_flaring_setup}

Some aspects of our setup itself still need to be investigated. To begin with, the flux variability of the flares is loosely constrained, for two main reasons. First, the flux variation is a measurement of how high the flare from the RS can peak over the FS. This is directly driven by the ratio in radiative efficiency between the FS and the RS, decided by the values of $\epsilon_e$ and $\epsilon_B$ in the FS, and the RS respectively. The timescale $\Delta t_\mathrm{obs} / t_\mathrm{obs}$ of the flare also depends on this ratio, as a better radiative efficiency of the FS will cause a shorter $\Delta t_\mathrm{obs}$. Improving the constraints on the microphysics of emission in the ejecta and the circumburst medium constitutes a crucial step towards validating the approach presented here. 

Second, the scope of this article focuses on cases in which the jet axis is aligned with the line of sight. We observe that, even though perturbations away from the line of sight give rise to flares with longer timescales than aligned perturbations, they also display smaller peak fluxes simply because of Doppler beaming effects. This gives rise to a range of values for $\Delta F / F$ for patches at arbitrary angles and can actually completely suppress the flare for high angles as the flux variability from the RS becomes unable to peak above the FS emission. As a result, we expect to mostly be seeing flares coming from perturbations aligned with the line of sight. 
More evidence is also being accumulated in favor of structured jets \citep[e.g.][]{Abbott2017,Troja2017a,Kasliwal2017,Kathirgamaraju2018b,Lazzati2019,Troja2019} with less energetic sides and domination by the central material, further advocating for more variability coming from the central region.
These aspects in turn favour the associated values of $\Delta t_\mathrm{obs} / t_\mathrm{obs} \sim 0.1 - 0.3$ corresponding to observations. 

These flares however can only fulfill the timescale requirements if the perturbed region remains sufficiently narrow. We observe that longer flares arise if the curvature effect starts to dominate the decay time, which can happen in the perturbed region becomes too wide. We find empirically that the timescale starts to increase for $\Delta \theta_c \gtrsim 0.01 \text{ rad}$. Actually, this constraint corresponds to the value of $1/\Gamma$ for $\Gamma = 100$ and is the expected size of angular perturbations due to transverse causal contact across the ejecta. As a result, we do not expect to see these longer flares, in accordance again with observations. This proves the robustness of this mechanism which does not rely on fine-tuning of the angle, nor the size of the patch in order to produce flares with the correct timescale, arrival time and flux variability.

Finally, this mechanism relies on the formation of a dense shell associated with radiating internal shocks in the ejecta, responsible for a first spike in the X-ray light-curve. One might wonder whether these internal shocks can directly contribute in gamma-rays. We do not expect that this is the case as the Lorentz factors involved in the tail of the ejecta are too small to shield the corresponding radiation from $\gamma\gamma$ annihilation. As a result, we expect these spikes not to be detectable in the prompt phase and to be only visible in cases where X-ray observations start very early, for which they can be interpreted as late prompt emission or early flares (such as some sometimes seen during the early steep decay phase). Our afterglow model thus does not depend on any prompt mechanism but merely requires the launch of a relativistic ejecta.


\section{Conclusions}

In this paper, we present one-dimensional SRHD numerical simulations of GRB afterglow blast waves. These simulations are carried out using our new moving-mesh code, improving both resolution and efficiency. We apply them to a novel scenario \citep{Hascoet2017,Lamberts2017a} for the origin of flares in the X-ray afterglow that involves the interaction of a long-lived reverse shock with a stratified ejecta. We implement numerical tracing of the boundaries of the local particle population and present the first broadband study of such scenario.

By exploring a larger parameter space than previous works, and by considering a more realistic description of the circumburst medium density, we are able to confirm that a chaotic shutdown of the central engine produces X-ray flares in the afterglow at an arbitrary long time after the burst.

In order to do so, the initial dynamics of the ejecta must include the following features:
\begin{itemize}
  \item A region (tail) of decreasing velocity as we move further towards the back of the ejecta, responsible for its expansion at it travels outwards, and delaying the interaction of a reverse shock with potential embedded features.
  \item Variations of the Lorentz factor in the tail. Each region of negative velocity gradient in this tail will be responsible for the formation of a dense shell as the faster material catches up with the slower material ahead, and later a flare when it is crossed by the RS. The number of flares in a single afterglow depends on the variability in the tail.
\end{itemize}

This model also relies on the assumption that the radiative efficiency of the forward shock needs to be smaller than that of the reverse shock, by means of a smaller value of $\epsilon_B$, such that flares from the RS can peak over the FS emission. Depending on the variability of the tail and the ratio of radiative efficiencies, a fraction of weak flares might go undetected.

The light curve displays two new features in comparison with a uniform profile. First, a spike is  produced by the internal shocks at early times, reminiscent of the mechanisms involved for prompt emission. The second feature is the result of the interaction of the RS with the dense region. The peak frequency and the cooling break of the RS both increase, leading to a flare in X-ray and a smaller bump in optical.

In order to retrieve the timescale flares expected from observations, we constrain the perturbation responsible for the flaring behavior to a small angular patch and show that the width of this patch actually naturally falls below the expected size derived by causal contact arguments as the cooling timescale needs to dominate over the curvature effect.

We show that this model produces flares at an arbitrarily long time after the burst if the perturbation in the gradual engine shutdown happens just before the complete end of its activity. This delays the interaction of the reverse shock with the resulting dense shell and produces flares with the same timescale and flux variability, but at later times.

Eventually, our moving mesh approach to the simulation of the dynamics is sufficiently efficient that we can expect to carry out additional simulations in a relatively short amount of time. Moving to two dimension would allow us to simulate the dynamics past jet spreading, and to study the link between flaring behavior and late evolution of the afterglow light-curve.

\section*{Acknowledgements}

We address special thanks to Geoffrey Ryan and Tanmoy Laskar for their respective help in setting up the moving-mesh code and the observational insight into GRB afterglow behavior that they provided.




\bibliographystyle{mnras}
\bibliography{biblio.bib} 




\appendix

\section{Code Validation} 
\label{sec:code_validation}

\begin{figure}
    \centering
    \includegraphics[width=0.45\textwidth]{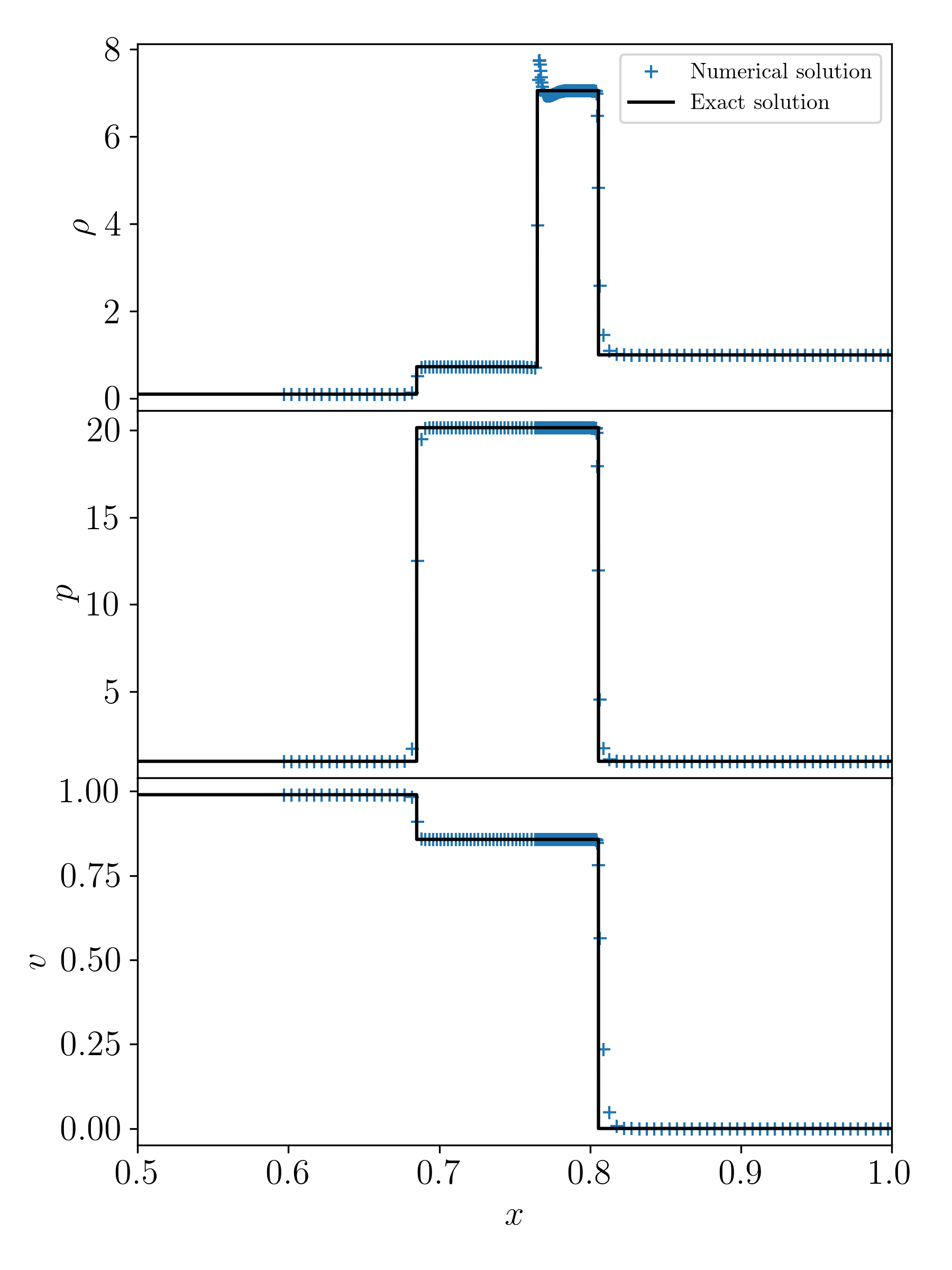}
    \caption{Shock tube in cartesian coordinates. $t=0.6$. Initial discontinuity at $x=0.25$. Left state: $\rho=0.1$, $p=1$, $v=0.99 \times c$. Right state: $\rho=1$, $p=1$, $v=0$. Initial resolution is 200 cells.}
    \label{fig:Shock-Tube_cart}
  \end{figure}

\begin{figure}
    \centering
    \includegraphics[width=0.45\textwidth]{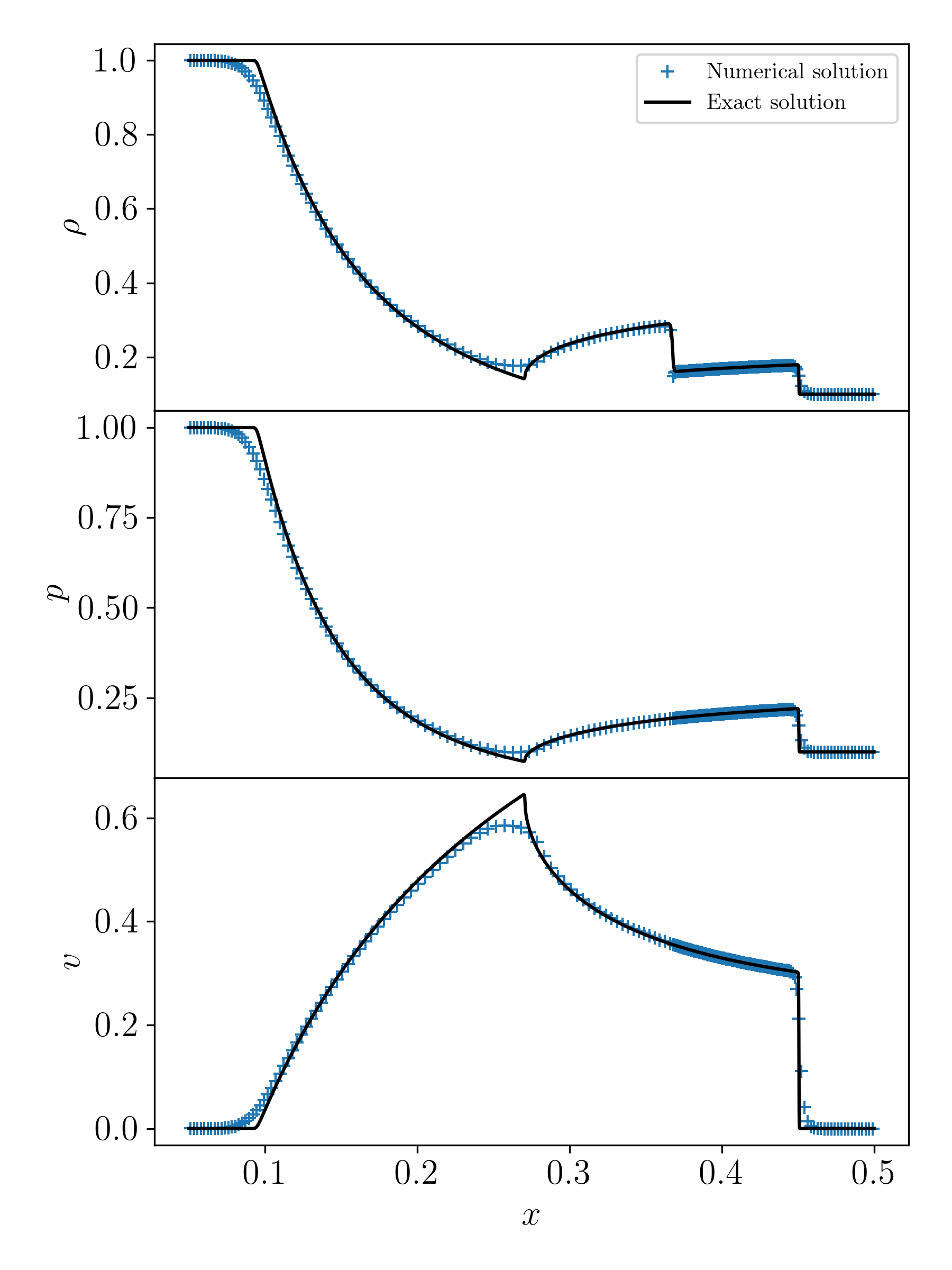}
    \caption{Shock tube in spherical coordinates. $t=0.3$. Initial discontinuity at $x=0.25$. Left state: $\rho=1$, $p=1$, $v=0$. Right state: $\rho=0.1$, $p=0.1$, $v=0$. Initial resolution is 200 cells.}
    \label{fig:Shock-Tube}
  \end{figure}

  First, the stability and accuracy of the code are tested on standard shock-tube setups. Fig. \ref{fig:Shock-Tube_cart} shows an example in cartesian coordinates in which the numerical solution is able to resolve the shock fronts and the contact discontinuity with high precision. The spike at the contact discontinuity is a resolution-dependent numerical oscillation linked to the initial transient state and is different from the physical density spike described in sec. \ref{sec:hydrodynamical_evolution}. Fig. \ref{fig:Shock-Tube} shows an example in spherical coordinates in which the numerical evolution is in good agreement with the exact solution. We tested the code on several other shock-tube setups that all yielded satisfying results but are not reported here. 

\begin{figure}
    \centering
    \includegraphics[width=0.45\textwidth]{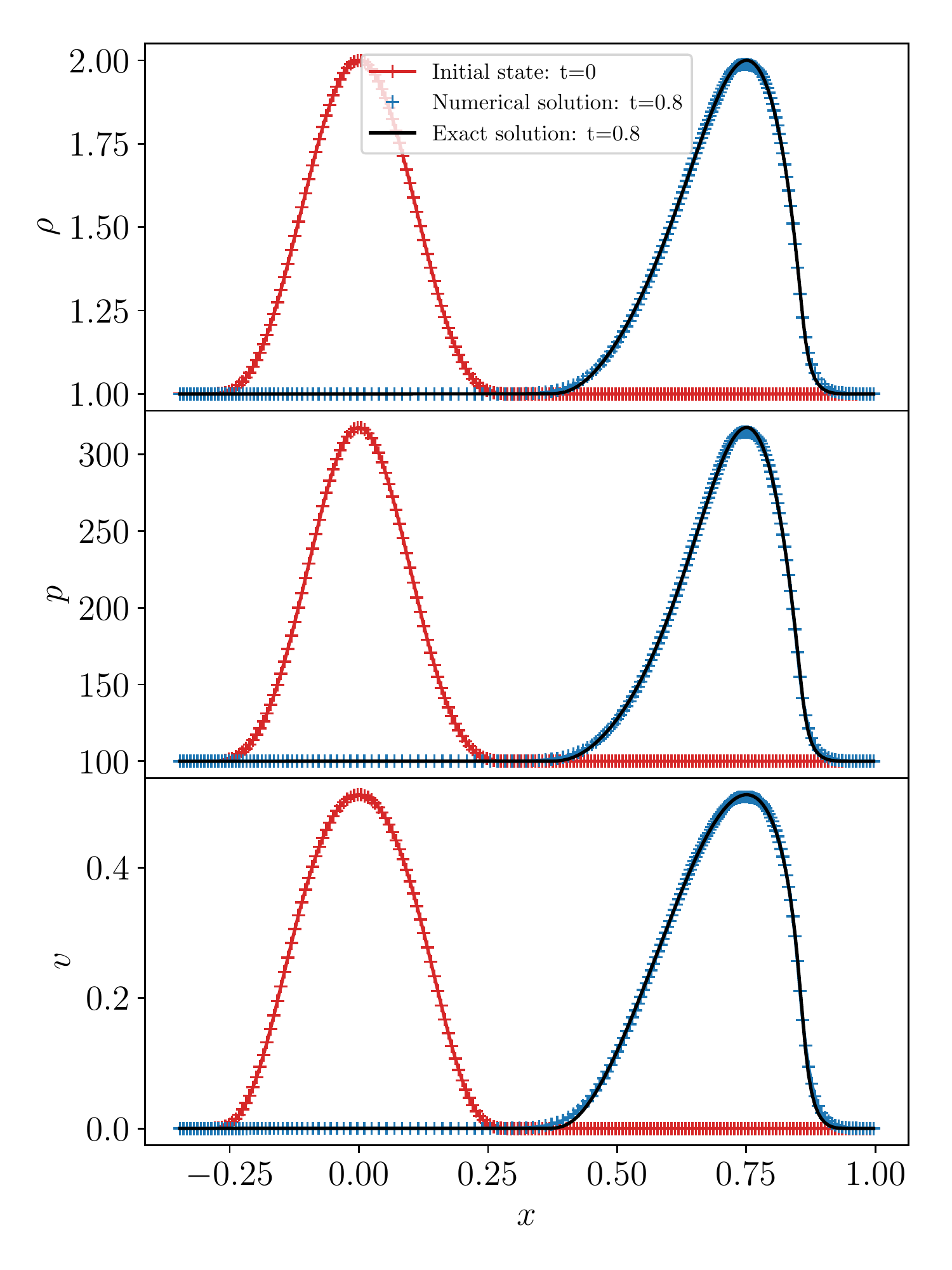}
    \caption{Isentropic wave test output for primitive variables. We plot the initial (left wave) and final state (right wave) on the same graph here. The numerical results at t=0.8 (blue crosses) are in very good agreement with the exact solution (black solid curve).}
    \label{fig:IsenWaveOutput}
  \end{figure}

\begin{figure}
    \centering
    \includegraphics[width=0.45\textwidth]{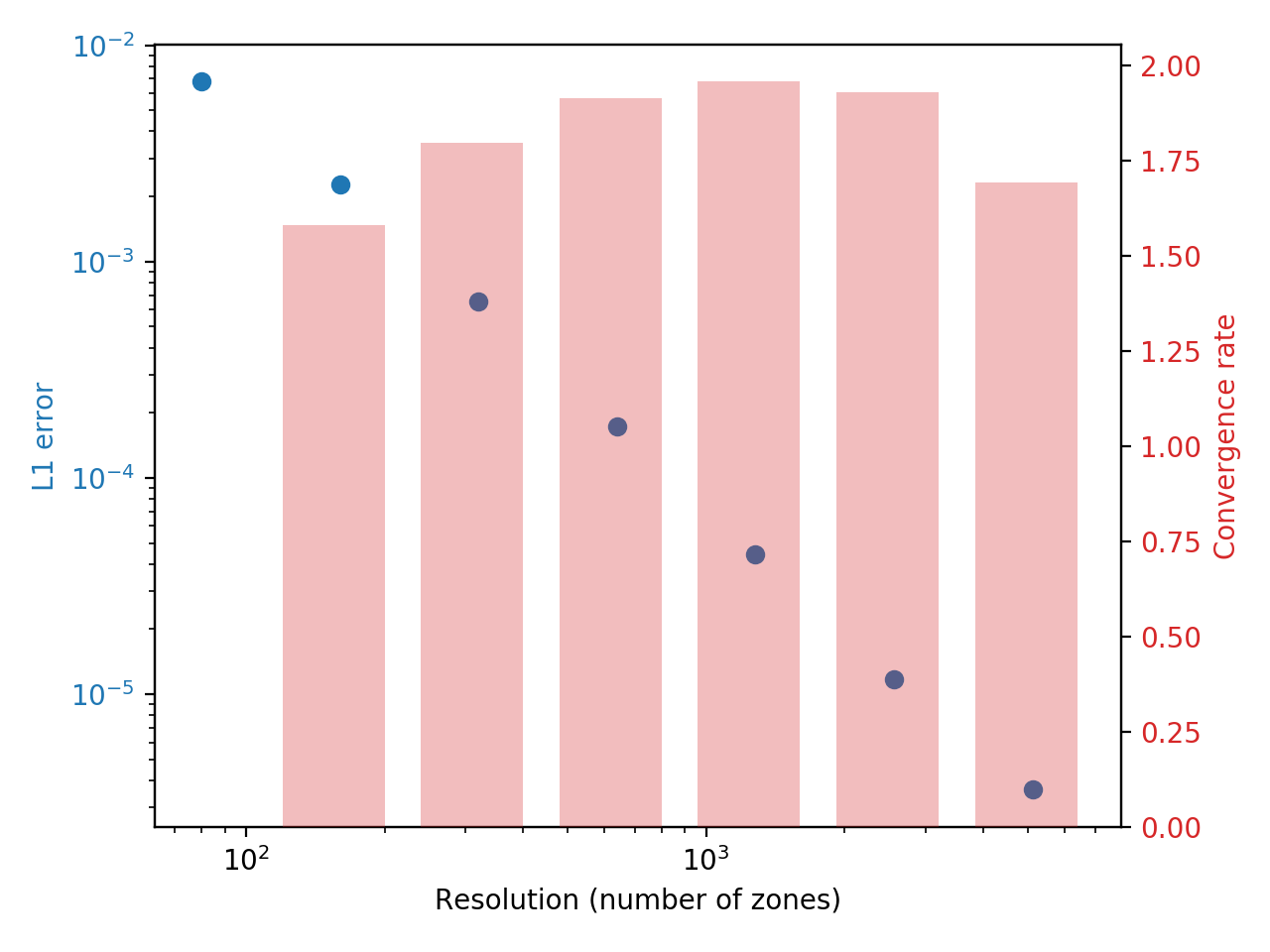}
    \caption{Convergence rate and L1 error for the isentropic wave test.}
    \label{fig:convOrderIsenWave}
  \end{figure}

\begin{figure}
    \centering
    \includegraphics[width=0.45\textwidth]{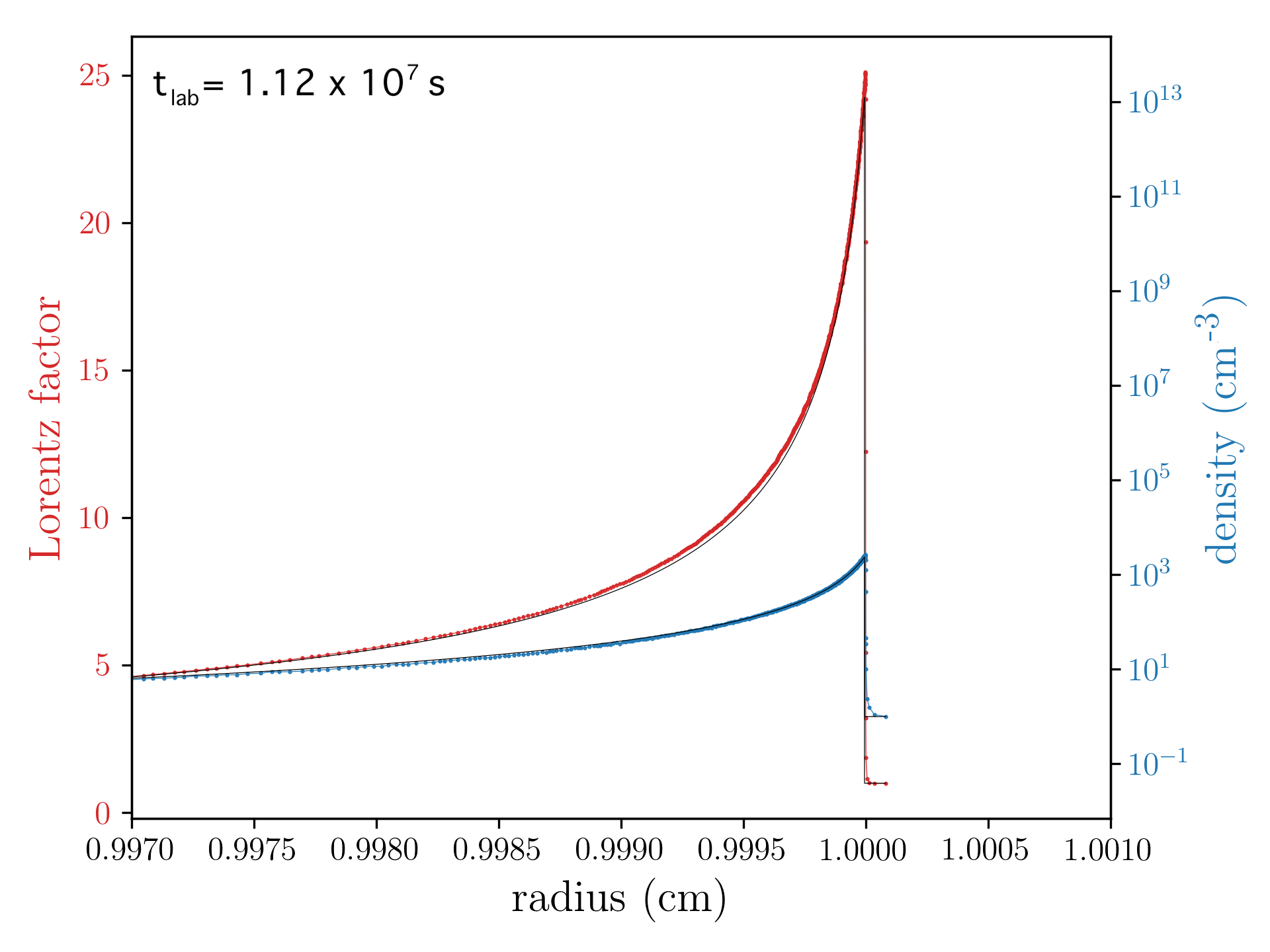}
    \caption{Validation of the moving-mesh code against the Blandford-McKee analytical evolution of a spherical blast-wave ($E_\mathrm{iso} = 10^{53}~\mathrm{erg}$, $n_0 = 1~\mathrm{cm}^{-3}$). Here we show the normalized density and Lorentz factor as a function of radius at the deceleration time. Then numerical solution (blue and red dots) is in very good agreement with the analytical Blandford-Mckee solution (solid black curve).}
    \label{fig:BM}
  \end{figure}

  We assess the accuracy of our code for smooth parts of the flow by testing the evolution of an isentropic wave in cartesian coordinates, following the setup presented in \citet{Zhang2006} and find that our code achieve the expected second order convergence in these cases, as is shown in figures \ref{fig:IsenWaveOutput} and \ref{fig:convOrderIsenWave}. Further testing is carried out by reproducing the Blandford-McKee (BM) setup \citep{Blandford1976} akin to the propagation of a spherically symmetric relativistic blast-wave in the ISM. The results are shown in fig. \ref{fig:BM} and show again that the code behaves as expected, even at significantly high Lorentz factors and with realistic values of circumburst medium number density $n_0 = 1 \text{ cm}^{-3}$. 
  
\section{Shock detection algorithm}
\label{sec:shock_detection_algorithm}

Shock detection methods usually rely on measuring gradients in the fluid state. However, these methods are only efficient for strong shocks and often fail to detect internal shock in our setups \citep{Lamberts2017a}. As a result, we follow the more robust method introduced in \citet{Rezzolla2003} and described in \citet{Zanotti2010}, that evaluates nature of the waves in the Riemann fan for all interfaces of the grid based on the calculation of limiting relative velocities. Indeed, in the absence of a magnetic field, a discontinuity between two regions of fluid with different thermodynamical values will decay into two waves propagating on both sides of a contact discontinuity (jump in density). These waves can either be rarefaction waves or shocks. As described above, we use this behavior to compute the dynamical evolution of our fluid by solving the Riemann problem at every interface between the cells of our grid. We can simply detect the shocks in our fluid by checking whether the wave pattern at each of these interfaces will contain a shock or not. A shock will appear if the relative velocity between neighbouring cells exceeds a threshold dependent on the fluid states.

The 1D aspect of our simulations allows us to compute the limiting relative velocities of each case of the Riemann problem rather efficiently. We report here the procedure presented in \citet{Rezzolla2003} to compute the limiting relative velocity in the Shock-Rarefaction ($\mathcal{SR}$) and 2-Shock ($2\mathcal{S}$) cases.

\paragraph*{One Shock - One Rarefaction wave case ($1\mathcal{R}_\leftarrow 3 \mathcal{C} 3^\prime \mathcal{S}_\rightarrow 2$)}

~\\
The Riemann fan produced at the interface between two neigbouring cells will contain at least a shock if the relative velocity between these cells $v_{12}$ is greater than a limiting velocity $(\tilde{v}_{12})_\mathcal{SR}$ given by
\begin{align}
  (\tilde{v}_{12})_\mathcal{SR} = \frac{1 - A_+(p_3)}{1 + A_+(p_3)} \bigg|_{p_3=p_2},
\end{align}
with 
\begin{align}
  A_+(p_3) \equiv \left[ \left( \frac{(\gamma - 1)^{1/2} - c_s(p_3)}{(\gamma - 1)^{1/2} + c_s(p_3)} \right) \left( \frac{(\gamma - 1)^{1/2} + c_s(p_1)}{(\gamma - 1)^{1/2} - c_s(p_1)} \right) \right]
    ^{2/(\gamma - 1)^{1/2}},
\end{align}
with $c_s(p_i)$ the sound speed in region $i$.

A very straightforward shock detection method would consist in comparing the relative velocities and limiting relative velocities at each interface, flagging a shock if:
\begin{align}
  v_{12} > (\tilde{v}_{12})_\mathcal{SR}.
\end{align}

However, this method detects shocks of any strength, and we are only interested in the strongest shocks responsible for particle acceleration. A nice workaround involves the calculation of the limiting relative velocities in the  $2\mathcal{S}$ case.

\paragraph*{Two-shock case ($1\mathcal{S}_\leftarrow 3 \mathcal{C} 3^\prime \mathcal{S}_\rightarrow 2$)} 
\label{par:two_shocks_case_}

  ~\\
  The fan will display two shocks if:
  \begin{align}
    v_{12} > (\tilde{v}_{12})_{2\mathcal{S}} = \sqrt{ \frac{(p_1 - p_2)(\hat{e} - e_2)}{(\hat{e} + p_2)(e_2 + p_1)}},
  \end{align}
  with
  \begin{align}
    \hat{e} = \hat{h} \frac{\gamma p_1}{(\gamma - 1)(\hat{h}-1)} - p_1,
  \end{align}
  where $\hat{h}$ is the only positive root of the Taub adiabat. In order to relax the our shock-detection criterion, we set the threshold for shock detection to a intermediate value between both relative velocity limits:
  \begin{align}
    (\tilde{v}_{12}) = (\tilde{v}_{12})_\mathcal{SR} + \chi[(\tilde{v}_{12})_{2\mathcal{S}} - (\tilde{v}_{12})_\mathcal{SR}],
  \end{align}
  where the value of $\chi$ is arbitrarily chosen between 0 and 1. In our simulations, choosing $\chi = 0.1$ gives satisfying results.



\bsp  
\label{lastpage}
\end{document}